\documentclass[journal]{IEEEtran}

\usepackage{color}
\usepackage{balance}

\usepackage{amssymb,amsfonts}
\usepackage{mathtools} 
\usepackage{amsmath, amsthm}

\usepackage{comment}
\usepackage{booktabs}
\usepackage{multirow}
\usepackage{tabularx}
\usepackage{array}
\usepackage{makecell}

\usepackage{graphicx}
\usepackage[font=footnotesize]{caption}
\usepackage{subcaption}
\usepackage{float}

\usepackage{geometry}

\usepackage{enumitem}

\usepackage{cite}
\usepackage[hidelinks]{hyperref}

\title{Coordinated Dynamic Operating Envelopes for Network-Admissible Flexibility at the Grid Edge}

\author{Ali Jalilian\textsuperscript{1},~\IEEEmembership{Student Member,~IEEE}, Deepjyoti Deka\textsuperscript{2},~\IEEEmembership{Senior Member,~IEEE}, Md. Umar Hashmi\textsuperscript{1},~\IEEEmembership{Senior Member,~IEEE}, and Dirk Van Hertem\textsuperscript{1},~\IEEEmembership{Fellow,~IEEE}
  \thanks{\textsuperscript{1} Division ELECTA of Departement Elektrotechniek (ESAT), KU Leuven University, Leuven, Belgium, and EnergyVille research center, Genk, Belgium.
  Emails: {\tt\small {\{ali.jalilian, mdumar.hashmi, dirk.vanhertem\}@kuleuven.be}}
  
  \textsuperscript{2} MIT Energy Initiative, MIT, Cambridge, Massachusetts, USA
  Email: {\tt\small {deepj87@mit.edu}}}
  }
  \vspace{-1em}

\begin{document}
\maketitle
\begin{abstract}
Dynamic operating envelopes (DOEs) provide a systematic framework to integrate the flexibility of distribution grid resources while safeguarding network limits such as line ratings and voltage bounds. However, the flexibility derived from individual DOEs is often restricted and conservative, especially when some resources can coordinate via communication with an aggregator. This paper presents a convex, geometry-aware framework for constructing DOE for distribution grid customers under partial coordination, with coordinated customers modeled through polytopal flexibility sets and non-coordinated customers through hyperrectangles. The framework additionally incorporates fairness constraints for export and import headroom allocated to the customers within the DOE design. To account for forecast uncertainty in inelastic injections, the DOE design is extended to a robust formulation for bounded uncertainty sets. Case studies on two European three-phase low-voltage feeders—a widely used test feeder and a large-scale (3589)-bus system—show that the proposed DOE construction expands aggregate flexibility while maintaining network feasibility, fairness, and robustness to forecast uncertainty. Coordinating 30\% of customers increases the aggregate active-power range by approximately 25\% on the test feeder, while coordinating 20 customers on the large-scale feeder increases it by approximately 43\%.
\end{abstract}



\section{Introduction}
\IEEEPARstart{T}{he} deployment of distributed energy resources (DERs)—including solar PV, batteries, and electric vehicles—has surged in recent years, driven by technological advances, declining costs, regulatory reforms, and consumer demand for clean energy~\cite{masson2024trends,iea2025global}. 
In parallel, flexibility markets, smart inverter capability mandates, and updated interconnection protocols such as IEEE 1547.2-2023 are being adopted by utilities and regulators worldwide to support safe and efficient DER integration~\cite{ieee1547.2-2023, iea2022der}. These developments reflect a multi-faceted push to enable clean energy participation and unlock DER value while respecting system safety requirements.

Traditional approaches such as static export limits, conservative connection studies, or fixed Volt/VAR settings often rely on worst-case assumptions and therefore leave substantial network headroom unused~\cite{liu2021grid, alam2023allocation}. Dynamic Operating Envelopes (DOEs), calculated either by distribution system operators (DSOs) or by intermediary operational platforms (e.g., DER management systems, or flexibility market platforms), provide a more adaptive alternative by communicating time-varying import/export limits that customers or aggregators can use in their local decisions to remain compatible with network constraints, while supporting increased DER utilization~\cite{liu2021project}.

More broadly, DOEs provide a practical interface for network-aware limits to be communicated to customers, aggregators, or markets without requiring the DSO to centrally control all downstream resources. This has motivated their use in
flexibility operation~\cite{hashminetwork}, hosting-capacity maximization~\cite{fani2024impact}, market design~\cite{alahmed2024decentralized}, and peer-to-peer trading~\cite{hoque2023dynamic}. Early DOE-like approaches were pioneered in Great Britain and Ireland, where non-firm connection schemes and flexible access pilots allowed renewable generators to connect under dynamic export limits that adjusted to network conditions~\cite{foote2013orkney,esb2021flexibleaccess}. Building on such precursors, Australia is widely regarded as a leading testbed for DOEs, where initiatives such as the \textit{``Accelerating the Implementation of Operating Envelopes''} project and Project EDGE have advanced DOE concepts from theory to practice~\cite{givisiez2024accelerating,aemo_projectedge}. Recent Australian distributor-led programs include flexible export and connection schemes by SA Power Networks, AusNet Services, and Energy Queensland, which dynamically adjust customer export limits according to available network capacity~\cite{SAPN_AusNet_FlexibleExports,AER_FlexibleExportLimits,EnergyQueenslandDynamicConnections}.

Most existing DOE designs communicate limits to individual prosumers, effectively treating each connection point independently~\cite{mahmoodi2023capacity,liu2023robust,hashmi2023robust}. This overlooks the growing role of coordinated energy communities and aggregators, which can improve hosting capacity, demand response, and economic efficiency while maintaining network security~\cite{BJARGHOV2024100154,scott2019network,jiang2024robust}. Among aggregation-based designs, approaches typically characterize a feasible region at a physical aggregation point~\cite{riaz2021modelling}, allocate headroom to aggregators anchored at predefined locations~\cite{jiang2025bargaining}, or model an aggregator-controlled portfolio while treating other injections as exogenous~\cite{song2025distribution}. In practice, however, participation in aggregators and flexibility platforms is  heterogeneous leading to coordinated and non-coordinated resources routinely coexisting on the same feeder. This leaves open a practical requirement that arises under
\emph{mixed} participation: a unified DOE framework that communicates a coupled feasible set to coordinated resources and individual envelopes to non-coordinated customers while preserving network feasibility.

A related but important distinction in DOE construction is between \emph{point feasibility} and \emph{set feasibility}. Methods that construct operating regions from finitely many OPF-derived boundary points~\cite{riaz2021modelling,lankeshwara2023time,yi2022fair,geth2025four} establish feasibility of those points, but not necessarily of every point within the communicated region. Other formulations explicitly construct envelopes contained in the network-feasible set~\cite{liu2023robust,mahmoodi2023capacity,gao2025equitable,russell2023robust}. We extend this set-feasibility requirement to partial coordination by ensuring that the complete allocation, i.e., the coupled DOE of coordinated participants together with individual envelopes for non-coordinated customers, is jointly feasible. Unlike independent operation, which naturally yields a hyperrectangular joint DOE, coordination permits richer coupled sets that capture trade-offs among participants. Constructing and enlarging such a set while jointly allocating individual envelopes is the central geometric problem addressed in Sections~\ref{sec:DOE model}-\ref{sec:poly_from_ellipsoid}.

Many DOE formulations communicate admissible flexibility only as active-power import/export ranges~\cite{alam2023allocation,liu2023robust,petrou2021ensuring,yi2022fair}. This focus reflects the central role of active power in customer energy transactions and billing, together with the practical simplicity of communicating and implementing scalar import/export limits. In these formulations, reactive power is either fixed according to an operating assumption or optimized internally as a control variable. Other works construct joint $P$--$Q$ operating regions, thereby communicating active and reactive power as separate flexibility dimensions~\cite{gerdroodbari2022dynamic,lankeshwara2023time,gao2025equitable}. Our proposed formulation uses the active-power range as its primary allocation objective. We first maximize the active-power DOEs while optimizing reactive-power injections as setpoints and then lift the resulting active-power DOEs to two-dimensional $P$--$Q$ operating regions. Further details are provided in Section~\ref{sec:pq_extension}.

A practical consideration for DOE deployment is the treatment of uncertainty in forecasts of demand and DER. Several envelope-construction methods are deterministic and do not embed an explicit uncertainty model in the DOE computation ~\cite{gao2025equitable,lankeshwara2023time}. This design choice relies on frequent re-computation to limit the impact of forecast errors. For example, \cite{petrou2021ensuring} adopts a 5-min update interval, and \cite{alam2023allocation} discusses practical mitigation strategies such as reserving headroom and imposing ramp-rate limits between successive envelopes. The authors in~\cite{mahmoodi2023capacity} handle uncertainty at the customer layer via affine recourse policies that keep the realized operating point within the assigned DOE. In contrast,~\cite{yi2022fair} formulates operating envelopes via a chance-constrained linear OPF, using Gaussian forecast-error assumptions. In our work, we incorporate forecast uncertainty directly in our DOE calculation via a standard budgeted robust formulation amenable to general distributions. Further we allow conservatism to be adjusted through a small number of interpretable parameters.

Finally, envelope design and allocation need to be principled and transparent, so as to minimize the disparity in allocations observed in distribution grid operations\cite{hashmi2022can}. A majority of fairness-aware DOE formulations \cite{azim2024dynamic,gao2025equitable}, however, are developed for the case in which all recipients are assigned the same type of resource and can therefore be compared on a common scalar basis. Related work has also distinguished the direction of flexibility by allocating export and import headroom separately~\cite{alam2023allocation}. Such allocation becomes insufficient under partial coordination as the coordinated group is represented by a coupled high-dimensional operating set, whereas non-coordinated customers require individual ranges. To overcome this, we introduce a tunable fairness constraint that reduces disparity in the allocation of maximum possible import and export flexibility across heterogeneous customers envelopes.

Table~\ref{tab:literature_comparison} compares representative DOE formulations according to five requirements relevant to the proposed framework: partial coordination, set feasibility, fairness, robustness to uncertainty in fixed loads, and the communication of joint $P$--$Q$ operating ranges.

\begin{table}[!t]
\centering
\footnotesize
\caption{Comparison of representative DOE formulations}
\label{tab:literature_comparison}
\vspace{-0.4em}
\renewcommand{\arraystretch}{1.15}
\setlength{\tabcolsep}{5pt}
\resizebox{\columnwidth}{!}{%
\begin{tabular}{lccccc}
\hline
Reference &
Partial coord. &
Set feas. &
Fairness &
Fixed-load unc. &
$P$--$Q$ range \\
\hline
\cite{yi2022fair}
& $\times$ & $\times$ & \checkmark & \checkmark & $\times$ \\
\cite{mahmoodi2023capacity,gao2025equitable}
& $\times$ & \checkmark & \checkmark & $\times$ & \checkmark \\
\cite{liu2023robust}
& $\times$ & \checkmark & \checkmark & $\times$ & $\times$ \\
\cite{lankeshwara2023time,riaz2021modelling}
& $\times$ & $\times$ & $\times$ & $\times$ & \checkmark \\
\cite{attarha2021network}
& $\times$ & $\times$ & $\times$ & $\times$ & $\times$ \\
\cite{alam2023allocation,petrou2021ensuring,Kumarawadu2025SmartMeter,jiang2025bargaining}
& $\times$ & $\times$ & \checkmark & $\times$ & $\times$ \\
\cite{gerdroodbari2022dynamic}
& $\times$ & \checkmark & $\times$ & $\times$ & \checkmark \\
\cite{geth2025four}
& $\times$ & $\times$ & \checkmark & $\times$ & \checkmark \\
\cite{russell2023robust}
& $\times$ & \checkmark & $\times$ & $\times$ & $\times$ \\
\cite{song2025distribution}
& $\times$ & \checkmark & $\times$ & \checkmark & \checkmark \\
\hline
\textbf{This paper}
& \checkmark & \checkmark & \checkmark & \checkmark & \checkmark \\
\hline
\end{tabular}
}
\end{table}

To summarize, this paper makes the following contributions:
\begin{itemize}[leftmargin=0pt,itemindent=*,align=left]
\item A DOE construction is formulated as a convex framework that accommodates partial coordination by assigning (i) a coupled $P$--$Q$ admissible region to a coordinated cohort and (ii) individual $P$--$Q$ DOEs to non-coordinated customers within a single network-feasible framework, maintaining feasibility for admissible combinations of their actions. These heterogeneous regions are jointly enlarged according to a common DOE objective, with feasibility enforced by containing the complete allocated set within the network-feasible injection region.

 \item A standard budgeted-uncertainty model based on \cite{bertsimas2004price} is incorporated into the DOE construction to account for uncertainty in inelastic injections, yielding envelopes that can be tuned for conservatism for general forecast-error distribution. 
 
 \item A tunable fairness constraint is introduced in the DOE construction to reduce disparity in the allocation of import and export flexibility across customers with heterogeneous envelope geometries, with quantification of impact on total flexibility. 
\end{itemize}
 
We validate the proposed framework on the European Low Voltage Test Feeder and a larger unbalanced three-phase feeder with 3589 nodes and 124 load points, where we also compare the resulting P--Q DOEs with an existing method~\cite{mahmoodi2023capacity}. On the European feeder, coordinating 30\% of customers expands the aggregate active-power injection range by approximately 25\%. On the larger feeder, coordinating 20 customers increases this range by approximately 43\% relative to the corresponding non-coordinated case and yields approximately three times the active-power range of the benchmark.

Our work thus highlights the substantial capability that prosumer coordination can play towards unlocking DER value and motivates the design of novel DOEs that leverage coordination effectively for system-level and prosumer gains.

The remainder of this paper is structured as follows. Section~\ref{sec:formulation} introduces the system model and presents our convex, geometry-aware framework for coordination-aware DOE design. Section~\ref{sec:extentions} extends the formulation to account for fixed-load uncertainty and to incorporate fairness constraints in the allocation of import/export headroom. Section~\ref{sec:case} reports numerical results and validates the proposed DOE construction under varying coordination levels as well as different robustness and fairness settings. Section~\ref{sec:conclusion} concludes with key insights.

\section{System Model for DOE design}
\label{sec:formulation}
In distribution grids, DOEs provide dynamic import/export limits for flexible customer injections to ensure that local device capabilities and system-level constraints, such as voltage and thermal, are not violated. We propose a novel geometry-aware, convex framework for generating DOEs that distinguishes two classes of participants:\\ (a) \textbf{Coordinated resources} (e.g., coordinated via an aggregator) with a joint DOE, and\\(b) \textbf{Non-coordinated resources}, which act independently and are given individual DOEs.

The DOE calculation is formulated as a DSO-side function. The DSO computes and approves admissible envelopes using the network model, current or forecast operating conditions, fixed-load forecasts, and capability ranges declared by customers or aggregators. These envelopes are implemented through existing institutional and contractual arrangements, including flexible connection agreements or aggregator contracts, or inverter/controller settings. Since individual and coordinated DOEs are computed jointly, compliant operation remains network safe. If operation outside an assigned DOE is detected, the DSO may apply existing remedies or contractual penalties, subject to the regulatory framework. The objective is to construct the largest admissible operating regions for flexible customers, both coordinated and non-coordinated. We first characterize the grid and power flow model, and describe its feasible injection set using operational constraints. We then introduce the geometric model for DOEs for different participant classes.

\subsection{Power flow model and Constraints}\label{sec:power flow}
We consider a radial distribution grid with $N+1$ nodes and $N$
lines. We use a linearized branch-flow model~\cite{kekatos2015voltage} to represent network constraints, as linear power flow models have been widely used for radial distribution feeders operation for optimization, control and estimation problems under typical LV conditions; the approximation derivation and accuracy are described in~\cite{baran2002optimal,gan2014convex,deka2017structure}. To verify feasibility, we evaluate constructed DOEs using nonlinear AC power-flow model in Section \ref{sec:adversarial test}.

Let $\mathbf{\hat{p}},\mathbf{\hat{q}}\in\mathbb{R}^N$ denote the vectors of active and reactive power injections at all nodes. Let $v_0$ denote the slack-bus squared voltage magnitude, $\mathbf{v}\in\mathbb{R}^N$ denote squared voltage magnitudes at all other nodes, and $\mathbf{P}\in\mathbb{R}^N, \mathbf{Q}\in\mathbb{R}^N$ collect active and reactive flows across all lines in the grid, respectively. Under the linearized branch flow model, we have:
 \begin{align}
 &\mathbf{v} \;=\; R \mathbf{\hat{p}} + X \mathbf{\hat{q}} + v_0 \mathbf{1}_N,
 \label{eq:volt_eq}\\
&\mathbf{P} = M\mathbf{\hat{p}}, ~~
 \mathbf{Q} = M\mathbf{\hat{q}},\label{eq:flow_eq}
 \end{align}
where $M\in \mathbb{R}^{N\times N}$ is the node to line reduced incidence matrix for the grid and $R,X \in \mathbb{R}^{N\times N}$ denote the sensitivity matrices mapping injections to nodal voltages. The formulation is three-phase: the vectors in \eqref{eq:volt_eq}--\eqref{eq:flow_eq} collect all energized node-phase quantities of the unbalanced feeder.

The operational limits on squared nodal voltage magnitudes are:
\begin{equation}
 \mathbf{v}^{\min} \;\leq\; \mathbf{v} \;\leq\; \mathbf{v}^{\max}.
 \label{eq:volt_bounds}
\end{equation}
Substituting \eqref{eq:volt_eq} into \eqref{eq:volt_bounds} yields \emph{voltage feasibility constraint}
\begin{equation}
 A^{(v)} \mathbf{\hat{p}} + B^{(v)} \mathbf{\hat{q}} \;\leq\; \mathbf{c}^{(v)}.
\end{equation}
where the upper and lower voltage limits are expressed as stacked linear inequalities. Similarly, active flow $P_\ell$ and reactive flow $Q_\ell$ on line $\ell$ with thermal rating $S_\ell^{\max}$ are constrained as 
 \begin{equation}
 P_\ell^2 + Q_\ell^2 \leq (S_\ell^{\max})^2, \quad \forall \ell.
 \label{eq:line thermal circle}
 \end{equation}
This defines a second-order cone in the $(P_\ell,Q_\ell)$ plane. To obtain a computationally tractable polyhedral description, we replace each circle \eqref{eq:line thermal circle} by a polygonal inner approximation~\cite{ahmadi2014linear}. Specifically, the circle is replaced with $2\rho$ half-spaces oriented at angles $\theta_r = \tfrac{\pi r}{\rho}$, $r=0, 1,\ldots,2\rho-1$, yielding linear \emph{line flow constraints} of the form
 \begin{equation}
 A^{(f)} \mathbf{\hat{p}} + B^{(f)} \mathbf{\hat{q}} \leq \mathbf{c}^{(f)}
 \label{eq:line flow approximation}
 \end{equation}
Note that by increasing $\rho$, the approximation can be tightened as needed. 

At each node $i$, we include local customer/device capabilities such as maximum power, inverter limits via a polygonal constraint on the injection $(p_i,q_i)$. Stacking these \emph{customer limits} for all buses yields
\begin{equation}
A^{(c)} \mathbf{\hat{p}} + B^{(c)} \mathbf{\hat{q}} \le \mathbf c^{(c)}.
\label{eq:poly_connection_limits_compact}
\end{equation}

\paragraph*{Feasible region for flexibility} 
By vertically stacking the constraint sets for voltages, line flow limits, and customer limits, we obtain the feasible operating space of injections $(\mathbf{\hat{p}},\mathbf{\hat{q}})$ as
\begin{subequations}
\begin{equation}
 \big\{ (\mathbf{\hat{p}},\mathbf{\hat{q}}) \in \mathbb{R}^{2N} \;\big|\;
 A \mathbf{\hat{p}} + B \mathbf{\hat{q}} \leq \mathbf{c} \big\}, \text{where~}
 \label{eq:polytope_final}
\end{equation}
\footnotesize
\begin{align}
 &A^T = \begin{bmatrix} {A^{(v)}}^T ~ {A^{(f)}}^T ~ {A^{(c)}}^T \end{bmatrix}, B^T = \begin{bmatrix} {B^{(v)}}^T ~ {B^{(f)}}^T ~ {B^{(c)}}^T \end{bmatrix},\nonumber\\
 &\mathbf{c}^T = \begin{bmatrix} {\mathbf{c}^{(v)}}^T ~ {\mathbf{c}^{(f)}}^T ~ {\mathbf{c}^{(c)}}^T \end{bmatrix} 
\end{align}\end{subequations}
\normalsize
We assume that each nodal injection comprises of a known fixed (non-flexible) component and a flexible component, i.e., 
\begin{equation}\label{eq:fixed_flex}
\mathbf{\hat{p}} = \mathbf{p}^{\text{fixed}}+\mathbf{p}, ~ \mathbf{\hat{q}} = \mathbf{q}^{\text{fixed}}+\mathbf{q}, ~~ {\mathbf s}^{\mathrm{fixed}} \triangleq [{{\mathbf p}^{\mathrm{fixed}}}^\top,{{\mathbf q}^{\mathrm{fixed}}}^\top]^\top.
\end{equation}
The DOE construction in this paper quantifies active-power flexibility. Accordingly, reactive power injections are optimized as setpoints rather than being included as DOE ranges. The feasible polytope \eqref{eq:polytope_final} for the flexible active power $\mathbf{p}$ can be represented as $\mathcal{H}_{\{\textbf{p}\}}$, where 
\begin{subequations}
\begin{align}
 &\mathcal{H}_{\{\textbf{p}\}}=\left\{ \mathbf{p} \in \mathbb{R}^N \;\middle|\; A \mathbf{p} \leq \mathbf{b}_\mathbf{q} \right\}, \\
 \text{with~~} &\mathbf{b}_\mathbf{q} = \mathbf{c}-H\mathbf{s}^{\text{fixed}}-B\mathbf{q}, 
 \text{~~where~~} H \triangleq [A ~B] \label{eq:bq def}
\end{align}
 \label{eq:total feasible region}
\end{subequations}
 Note that the flexibility region is centered around the fixed operating point. Next, we define the mathematical models for individual and coordinated DOEs, and then develop an optimization problem to maximize the system flexibility.

\subsection{Modeling coordinated and non-coordinated DOEs}\label{sec:DOE model}
We consider two categories of customers, coordinated and non-coordinated. Let $\mathcal{N}$ denote the set of $n_{\mathcal{N}}$ customers that are non-coordinated and have independent flexible operating points $\textbf{p}^\mathcal{N}$. This is similar to the existing setting in \cite{liu2023robust, russell2023robust}. Their feasibility region, $\mathcal{H}_{\{\textbf{p}^\mathcal{N}\}}$, is expressed as a hyperrectangle (axis-aligned box) given by:
\begin{equation}
 \mathcal{H}_{\{\mathbf{p}^{\mathcal{N}}\}} = \left\{ \mathbf{p}^{\mathcal{N}} \in \mathbb{R}^{n_{\mathcal{N}}} \; \middle| \; P_i^- \leq p_i \leq P_i^+, \; \forall i \in \mathcal{N} \right\} \label{eq:non-com. feas. s}
\end{equation}
The geometric rationale is direct, as a hyperrectangle enables an admissible interval for each non-coordinated customer regardless of flexible injections elsewhere. This independence guarantee is also the source of its conservatism: because the DSO cannot condition one non-coordinated customer's admissible action on the simultaneous actions of others, the entire Cartesian product of the assigned intervals must remain feasible. The total volume of this region is:
\begin{equation}\label{eq:vol_uncor}
 \text{V}_{\mathcal{N}} = \prod_{i \in \mathcal{N}} (P_i^+ - P_i^-)
\end{equation}

\begin{figure}[ht!]
 \centering
 \includegraphics[clip, trim=3cm 3cm 3cm 2.5cm, width=0.65\linewidth]{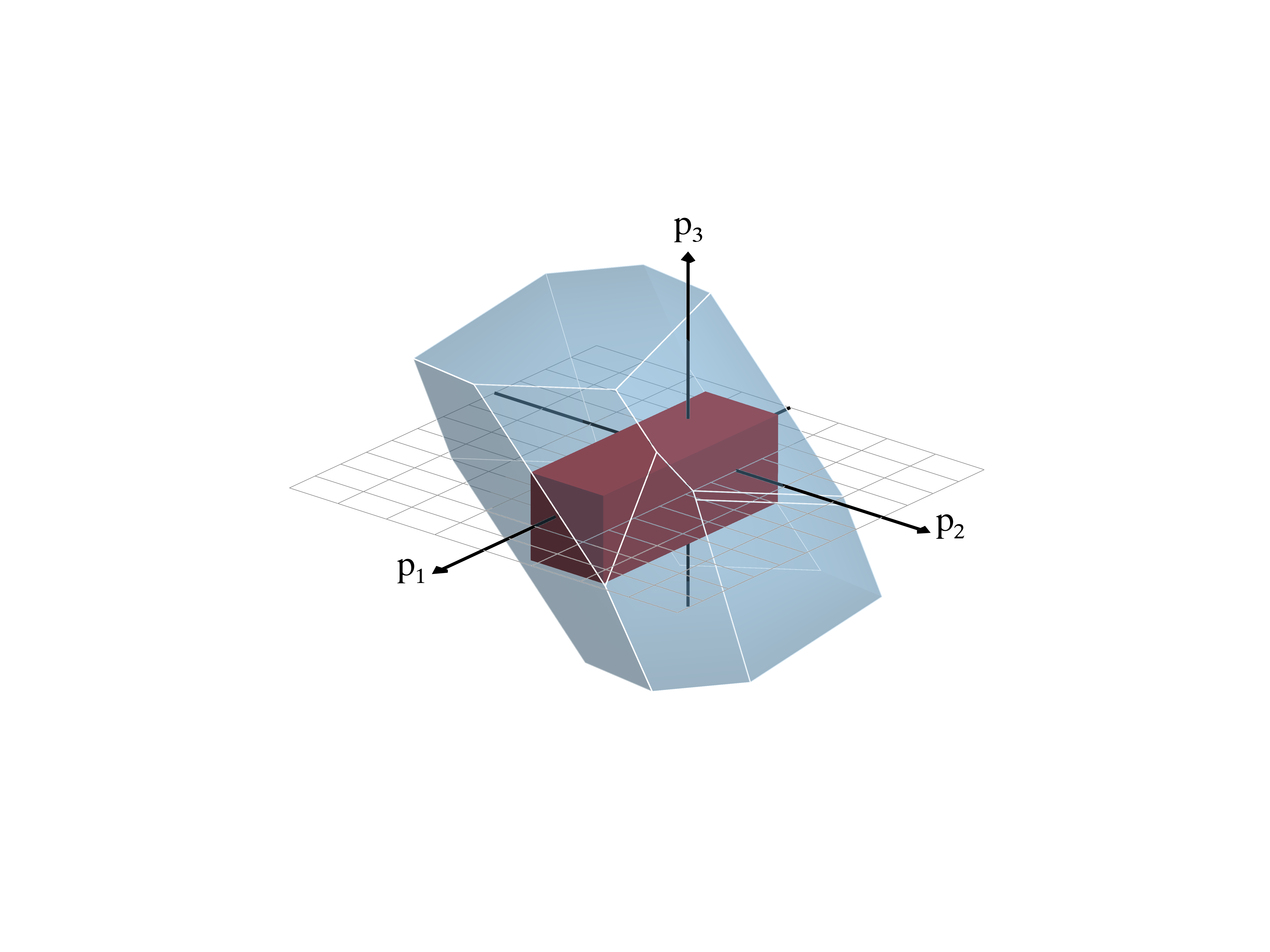}
 \caption{Coordinated (polyhedral) DOE versus independence-limited (axis-aligned) DOE. The hyperrectangle must fit inside the polytope to remain feasible for all uncoordinated choices.}
 \label{fig:DOE poly vs hyper}
\end{figure}

Next, let $\mathcal{M}$ be the set of $n_{\mathcal{M}}$ $=(N-n_{\mathcal{N}})$ coordinated customers, such that their flexible injection $\textbf{p}^\mathcal{M}$ can be managed jointly. Their aggregated flexibility region, denoted $\mathcal{H}_{\{\mathbf{p}^{\mathcal{M}}\}}$, can have an arbitrary shape with adjustments across participants, such that the feasibility constraints Eq.~\ref{eq:total feasible region} are satisfied. Figure~\ref{fig:DOE poly vs hyper} illustrates this contrast for a set of three customers: the polyhedral DOE achievable with coordination is much larger than the largest axis-aligned box that can be offered without it.

To obtain a tractable, geometry-aware objective for the volume of coordinated DOE, we employ an inscribed ellipsoidal representation as a proxy. Ellipsoids admit a compact parametrization via a centre and a positive semidefinite (PSD) shape matrix, and enable direct maximization within a convex program. After solving the design problem, we re-construct the DOE for the coordinated group as a polytope as explained in Section \ref{sec:poly_from_ellipsoid}.

The ellipsoidal parametric form for $\mathcal{H}_{\{\mathbf{p}^{\mathcal{M}}\}}$ defined as \cite{boyd2004convex}:
\begin{equation}
 \mathcal{H}_{\{\mathbf{p}^{\mathcal{M}}\}} 
 = \left\{ \mathbf{p}^{\mathcal{M}} \in \mathbb{R}^{n_{\mathcal{M}}} \;\middle|\;
 \mathbf{p}^{\mathcal{M}} = W \mathbf{u} + \mathbf{g}, \; \|\mathbf{u}\|_2 \leq 1 \right\} \label{eq:com. feas. s}
\end{equation}
Here, $\mathbf{g} \in \mathbb{R}^{n_{\mathcal{M}}}$ is the center of the ellipsoid and $W \in \mathbb{R}^{n_{\mathcal{M}} \times n_{\mathcal{M}}}$ is a symmetric PSD matrix that defines the shape and orientation of the ellipsoid. $\textbf{u} \in \mathbb{R}^{n_{\mathcal{M}}}$ is an auxiliary vector constrained to lie within the unit Euclidean ball. The volume of the ellipsoid defined above is given by:
\begin{equation}\label{eq:vol_cor}
 \text{V}_{\mathcal{M}} = \mathrm{Vol}(\mathbb{B}^{n_{\mathcal M}}) \cdot \det(W)= \frac{\pi^{n_{\mathcal{M}}/2}}{\Gamma\left(n_{\mathcal{M}}/2 + 1\right)} \cdot \det(W),
\end{equation}
where $\mathrm{Vol}(\mathbb{B}^{n_{\mathcal M}})$ is the volume of the unit $n_{\mathcal M}$-ball. $\Gamma(\cdot)$ is the Gamma function. $\det(W)$ determines the scaling of the ellipsoid's volume; a larger $\det(W)$ corresponds to more coordinated DOE space.

\subsection{Maximizing DOE}\label{sec: DOE maximization}
We now seek to compute the largest aggregate DOE that remains feasible for the underlying grid constraints. Using \eqref{eq:vol_uncor} and \eqref{eq:vol_cor}, the volume of the aggregate DOE—given by the Cartesian product of the non-coordinated and coordinated sets— is the product of $ \text{V}_{\mathcal{N}}$ for the hyperrectangle $\mathcal{H}_{\{\mathbf{p}^{\mathcal{N}}\}}$ and $\text{V}_{\mathcal{M}}$ for the ellipsoidal set $\mathcal{H}_{\{\mathbf{p}^{\mathcal{M}}\}}$. The constant term $\frac{\pi^{n_{\mathcal{M}}/2}}{\Gamma(n_{\mathcal{M}}/2 + 1)}$ in \eqref{eq:vol_cor} is independent of the optimization variables and can be safely omitted from the objective. We apply a logarithmic transformation to the objective that preserves the ordering but transforms the product into a sum:

\begin{equation}
 \max \quad \log \det(W) + \sum_{i \in \mathcal{N}} \log (P_i^+ - P_i^-)
 \label{eq:objective log}
\end{equation}
This transformation yields a concave objective function because $\log \det(W)$ is concave over the domain of symmetric positive definite matrices $W \succ 0$, and $\log(P_i^+ - P_i^-)$ is concave for all $P_i^+ > P_i^-$. 

\paragraph*{Feasibility} To ensure that DOEs are admissible, we need to ensure that the system feasibility constraint \eqref{eq:total feasible region} is satisfied for any combination of flexible injections from non-coordinated $\mathbf{p}^{\mathcal{N}}$ and coordinated $\mathbf{p}^{\mathcal{M}}$, given in \eqref{eq:non-com. feas. s} and \eqref{eq:com. feas. s}, respectively. 
\begin{equation}
 A \begin{bmatrix} \mathbf{p}^{\mathcal{M}} \\ \mathbf{p}^{\mathcal{N}} \end{bmatrix} \leq \mathbf{b}_\mathbf{q}, \quad \forall \mathbf{p}^{\mathcal{M}} \in \mathcal{H}_{\{\mathbf{p}^{\mathcal{M}}\}}, \quad\forall\mathbf{p}^{\mathcal{N}} \in 
 \mathcal{H}_{\{\mathbf{p}^{\mathcal{N}}\}} 
  \label{eq:total feas set} 
\end{equation}

Decomposing $A = [A_{\mathcal{M}} \;\; A_{\mathcal{N}}]$ along its columns between $\mathcal{M}$ and $\mathcal{N}$ in \eqref{eq:total feas set}, we have
\begin{equation}
 A_{\mathcal{M}} (W \mathbf{u} + \mathbf{g}) + A_{\mathcal{N}} \mathbf{p}^{\mathcal{N}} \leq \mathbf{b}_\mathbf{q}; 
  \forall \|\mathbf{u}\|_2 \leq 1, \mathbf{p}^{\mathcal{N}} \in [P^-_i, P^+_i]
  \label{eq:before_robust}
\end{equation}
To ensure that this inequality holds for all admissible $\mathbf{u}$ and $\mathbf{p}^{\mathcal{N}}$, we bound the worst-case value as follows:

\begin{itemize}[leftmargin=0pt,itemindent=*,align=left]
 \item The maximum of $A_{\mathcal{M}} W \mathbf{u}$ over $\|\mathbf{u}\|_2 \leq 1$ is given by the operator norm $\|A_{\mathcal{M}} W\|_2$.
 \item The worst-case contribution from $A_{\mathcal{N}} \mathbf{p}^{\mathcal{N}}$ is computed by splitting the matrix $A_{\mathcal{N}}$ into its positive and negative parts, $A_{\mathcal{N}} = A_{\mathcal{N}}^+ + A_{\mathcal{N}}^-, $ where $(A_{\mathcal{N}}^+)_{ij} = \max\{(A_{\mathcal{N}})_{ij}, 0 \}$, and $(A_{\mathcal{N}}^-)_{ij} = \min\{(A_{\mathcal{N}})_{ij}, 0 \}$. Then, the maximum is $A_{\mathcal{N}}^+ \mathbf{P}^+_{\mathcal{N}} + A_{\mathcal{N}}^- \mathbf{P}^-_{\mathcal{N}}$.
\end{itemize}

Putting it all together, we obtain the following sufficient constraint:
\begin{equation}
 \|A_{\mathcal{M}} W\|_2 + A_{\mathcal{M}} \mathbf{g} + A_{\mathcal{N}}^+ \mathbf{P}^+_{\mathcal{N}} + A_{\mathcal{N}}^- \mathbf{P}^-_{\mathcal{N}} \leq \mathbf{b}_\mathbf{q},
 \label{eq:main inequality}
\end{equation}
where $\mathbf{b}_\mathbf{q}$ is given in \eqref{eq:bq def}. This ensures that the entire ellipsoid $\mathbf{p}^{\mathcal{M}}$ and hyperrectangle $\mathbf{p}^{\mathcal{N}}$ lie within the feasible region.

Finally, we need to ensure that the zero-injection point lies within the DOEs, as this represents the baseline operating condition where no flexibility is activated. For non-coordinated customers, it suffices to enforce that the origin belongs to the hyperrectangle, i.e.,
\begin{equation}
 P_i^- \leq 0 \leq P_i^+, \quad \forall i \in \mathcal{N},
 \label{eq:zero_non}
\end{equation}
For coordinated customers, the joint DOE must contain the origin after aggregating with the contributions of the non-coordinated customers. This condition is satisfied if the terms corresponding to the coordinated customers vanish in the constraint \eqref{eq:main inequality}. Accordingly, we enforce the condition:
\begin{align}
 A_{\mathcal{N}}^+ \mathbf{P}^+_{\mathcal{N}} + A_{\mathcal{N}}^- \mathbf{P}^-_{\mathcal{N}} \leq \mathbf{b}_\mathbf{q}
 \label{eq:zero_coord}
\end{align}

\textbf{Preliminary DOE design:} We can now formulate the DOE design problem as a convex program. The variables are the non-coordinated bounds $\{P_i^-,P_i^+\}_{i\in\mathcal N}$, the coordinated ellipsoid parameters $(W,\mathbf g)$, and the reactive-power setpoints $\mathbf q$. The objective maximizes the aggregate DOE. Feasibility is enforced by requiring that all combinations $\mathbf p^{\mathcal M}\!\in\!\mathcal H_{\{\mathbf p^{\mathcal M}\}}$ and $\mathbf p^{\mathcal N}\!\in\!\mathcal H_{\{\mathbf p^{\mathcal N}\}}$ satisfy \eqref{eq:total feasible region}, using the sufficient robust constraint \eqref{eq:main inequality}, together with \eqref{eq:zero_non}--\eqref{eq:zero_coord} to keep the baseline admissible.

\begin{subequations}
\begin{align}
 \max_{\substack{
P^\pm_i, \mathbf{q},W, \mathbf{c} 
}}
\quad 
 & \log \det(W) + \sum_{i \in \mathcal{N}} \log (P_i^+ - P_i^-) \\[6pt]
 \text{s.t.} \quad 
 & \eqref{eq:bq def}, \eqref{eq:main inequality}, \eqref{eq:zero_non}, \eqref{eq:zero_coord}\\ 
 & W \succeq 0.
\end{align}
\label{eq:main surr. prob}
\end{subequations}

The objective in \eqref{eq:main surr. prob} specifies the DSO's geometric criterion for enlarging the coordinated DOE within the network-feasible region. The log-det term favors a residual feasible region that contains a large inner set with extent in multiple directions, rather than increasing flexibility only along selected operating directions. If service-specific price signal or dispatch priority is available, the same set-containment constraints can be retained and combined with alternative objectives, such as weighted import/export margins or price-weighted flexibility measures.

\subsection{Construction of the final polytope}
\label{sec:poly_from_ellipsoid}
After solving \eqref{eq:main surr. prob}, let \({\mathbf q}^*_{\mathcal N}\), \({\mathbf q}^*_{\mathcal M}\), \(\{ P_i^{-*}, P_i^{+*}\}_{i\in\mathcal N}\), and \((W^*,{\mathbf c}^*)\) denote the optimized reactive-power setpoints, the active-power limits for the hyperrectangular DOE of non-coordinated customers, and the coordinated ellipsoid parameters, respectively. The aggregate DOE of the coordinated customers is obtained as a polytope, by subtracting the worst-case contributions of the non-coordinated customers $\mathcal{N}$ from the total feasible set in \eqref{eq:total feas set}. Using \eqref{eq:bq def}, the polytopal DOE for the coordinated customers is expressed as:
\begin{align}
 &\mathcal{P}^*_{\{\mathbf{p}^{\mathcal{M}}\}} = \left\{ \mathbf{p}^{\mathcal{M}} \in \mathbb{R}^{n_{\mathcal{M}}} \;\middle|\; 
 A_{\mathcal{M}} \mathbf{p}^{\mathcal{M}} \leq \mathbf{b}^{\mathrm{res}} \right\}, \text{~where} \label{eq:final polytope}\\
 &\mathbf{b}^{\mathrm{res}} = \mathbf{c}-H\mathbf{s}^{\text{fixed}}-B\mathbf{q}^* - \hspace{-2pt} 
 \sum_{i \in \mathcal{N}}\hspace{-2pt} \left( A_{\mathcal{N}_i}^+ {P}_i^{+*} + A_{\mathcal{N}_i}^- {P}_i^{-*} \right).
\label{eq:residual rhs}
\end{align}

It is important to note that the ellipsoid in \eqref{eq:com. feas. s} is used only inside the optimization problem as a tractable inner-size surrogate. The log-det term maximizes the volume of an ellipsoid contained in the residual coordinated feasible region, which encourages multidirectional flexibility and avoids solutions that expand only in a few directions while preserving convexity of the design problem.

\subsection{Extension to active--reactive DOEs}
\label{sec:pq_extension}

The DOE construction above publishes active-power regions, while the reactive-power vector is optimized as a setpoint through \eqref{eq:bq def} and \eqref{eq:main surr. prob}. This modeling choice reflects the fact that many practical DOE and distribution-level pricing mechanisms primarily value active-power import/export capacity. Since reactive power can be a valuable controllable degree of freedom for inverter-based DERs, we now present how the active power hyperrectangle and ellipsoidal solution given by Problem \eqref{eq:main surr. prob} can be lifted to form active--reactive DOE regions.

Consider the solution ${\mathbf q}^*_{\mathcal N}, {\mathbf q}^*_{\mathcal M}$, $\{P_i^{-*},P_i^{+*}\}_{i\in\mathcal N}$ and $(W^*,{\mathbf g}^*)$ for Problem \eqref{eq:main surr. prob}. Note that the reactive power setpoints and active power ranges satisfy constraint \eqref{eq:main inequality}.

We first evaluate how much constraint margin is occupied by the optimized active-power hyperrectangle and ellipsoid at the optimized reactive-power setpoint. Using $B \triangleq [B_{\mathcal M}~B_{\mathcal N}]$, we define:
\begin{subequations}
\begin{align}
&\underline{\boldsymbol{\beta}}^{\mathcal N}_i
\triangleq
A_{\mathcal N_i}^{+}\hat P_i^{+*}
+
A_{\mathcal N_i}^{-}\hat P_i^{-*}
+
B_{\mathcal N_i}q^*_i,\quad \forall i\in\mathcal N,\label{eq:pq_preliminary_noncomm_budget}\\
&\underline{\boldsymbol{\beta}}^{\mathcal M} 
\triangleq
\left\|A_{\mathcal M}W^*\right\|_2
+
A_{\mathcal M}{\mathbf g}^*
+
B_{\mathcal M}{\mathbf q}^*_{\mathcal M},
\label{eq:pq_preliminary_coordinated_budget}
\end{align}
\end{subequations}

From \eqref{eq:main inequality}, \eqref{eq:bq def}, these preliminary budgets satisfy $\underline{\boldsymbol{\beta}}^{\mathcal M}
+
\sum_{i\in\mathcal N}
\underline{\boldsymbol{\beta}}^{\mathcal N}_i
\leq
\mathbf c-H\mathbf s^{\mathrm{fixed}}$.
The remaining row-wise capacity is therefore
\begin{equation}
\mathbf r
=
\mathbf c-H\mathbf s^{\mathrm{fixed}}
-
\underline{\boldsymbol{\beta}}^{\mathcal M}
-
\sum_{i\in\mathcal N}
\underline{\boldsymbol{\beta}}^{\mathcal N}_i
\geq
\mathbf 0 .
\label{eq:pq_residual_capacity}
\end{equation}
We assign this residual capacity among \(P\)--\(Q\) DOEs while preserving feasibility. For rows with nonzero preliminary contribution, this is done using the following contribution rule:
\begin{subequations}
\begin{align}
&\boldsymbol{\delta}^{\mathcal N}_i
=
\mathbf r
\cdot
\frac{
\left|\underline{\boldsymbol{\beta}}^{\mathcal N}_i\right|
}{
\mathbf s_{\beta}
},
~\forall i\in\mathcal N,\quad\text{and}
\quad
\boldsymbol{\delta}_{\mathcal M}
=
\mathbf r
\cdot
\frac{
\left|\underline{\boldsymbol{\beta}}^{\mathcal M}\right|
}{
\mathbf s_{\beta}
},
\label{eq:pq_coordinated_inflation} \\
&\text{where~~} \mathbf s_{\beta}
=
\left|
\underline{\boldsymbol{\beta}}^{\mathcal M}
\right|
+
\sum_{i\in\mathcal N}
\left|
\underline{\boldsymbol{\beta}}^{\mathcal N}_i
\right|.
\label{eq:pq_absolute_contribution}
\end{align}    
\end{subequations}
The active--reactive DOEs assigned to non-coordinated customers in $\mathcal{N}$ and coordinated set $\mathcal{M}$ are given by 
\begin{equation}
\begin{split}
&\mathcal D_i^{PQ}
=\hspace{-1pt}
\left\{
(p_i,q_i)\in\mathbb R^2
\;\middle|\;\hspace{-1pt}
A_{\mathcal N_i}p_i
+
B_{\mathcal N_i}q_i
\leq
\underline{\boldsymbol{\beta}}^{\mathcal N}_i
+
\boldsymbol{\delta}^{\mathcal N}_i
\right\},\\
&\mathcal P_{\mathcal M}^{PQ}
=
\left\{\begin{bmatrix}
\mathbf p^{\mathcal M}\\
\mathbf q^{\mathcal M}
\end{bmatrix}
\in\mathbb R^{2n_{\mathcal M}}
\;\middle|\;
\begin{bmatrix}
A_{\mathcal M} & B_{\mathcal M}
\end{bmatrix}
\begin{bmatrix}
\mathbf p^{\mathcal M}\\
\mathbf q^{\mathcal M}
\end{bmatrix}
\leq
\underline{\boldsymbol{\beta}}^{\mathcal M}
+
\boldsymbol{\delta}_{\mathcal M}
\right\}
\end{split}
\label{eq:coordinated_pq_doe}
\end{equation}
By construction, the assigned margins use the full admissible constraint margin and stays feasible. Furthermore, the original active-power intervals and coordinated ellipsoid are preserved at the optimized reactive-power slices in $\mathcal D_i^{PQ}$ and $\mathcal P_{\mathcal M}^{PQ}$, respectively. Note that the margin-expansion rule in \eqref{eq:pq_coordinated_inflation} is one feasible way to lift the optimized active-power DOE to an active--reactive DOE. Other expansion rules can be used provided that the final margins dominate the preliminary margins.

\section{Robustness and Fairness Extensions}
\label{sec:extentions}

\subsection{Fixed-Load Uncertainty}
\label{subsec:robust_fixed_load}
The proposed DOE construction in \eqref{eq:main surr. prob} relies on a decomposition of net injections into fixed and flexible components as given in \eqref{eq:fixed_flex}. In practice, the fixed component (e.g., non-controllable demand) is subject to forecasting error. To ensure network feasibility under uncertainty in fixed load $\mathbf s^{\mathrm{fixed}}$, we adopt a tunable bounded-support budgeted uncertainty model of Bertsimas--Sim \cite{bertsimas2004price} within our DOE design.

We model $\mathbf s^{\mathrm{fixed}}$ in \eqref{eq:fixed_flex} as sum of known $\bar{\mathbf s}^{\mathrm{fixed}}$ and uncertainty part:
\begin{equation}
\mathbf s^{\mathrm{fixed}} = \bar{\mathbf s}^{\mathrm{fixed}} + \Delta \boldsymbol{\zeta},
\label{eq:fixed_load_uncertainty_model}
\end{equation}
where $\boldsymbol{\zeta}\in\mathbb{R}^{N}$ is an uncertainty vector scaled by component-wise known magnitudes $\Delta=\mathrm{diag}(\Delta_1,\dots,\Delta_{N})\succeq 0$. The budgeted box uncertainty set for $\boldsymbol{\zeta}$ is given as:
\begin{equation}
\mathcal U(\Gamma) \triangleq \left\{\boldsymbol{\zeta}\in\mathbb{R}^{N}:\ \|\boldsymbol{\zeta}\|_\infty \le 1,\ \|\boldsymbol{\zeta}\|_1 \le \Gamma \right\}.
\label{eq:budgeted_box_set}
\end{equation}
Here $\Gamma\in[0,N]$ controls conservatism: $\Gamma=0$ recovers the nominal case with no uncertainty, while $\Gamma=N$ corresponds to full worst-case box uncertainty.

The uncertainty parameters have direct operational interpretations: \(\Delta_i\) bounds the fixed-load forecast error at component \(i\), while \(\Gamma\) controls the allowed simultaneity of deviations across components. These margins can be informed by historical forecast-error data, for example by selecting \(\Delta_i\) from empirical absolute-error quantiles for relevant feeder, customer-class, or time-of-day groups, and tuning \(\Gamma\) to the desired protection level.

Recall the DOE feasibility constraint \eqref{eq:main inequality} with $\mathbf{b}_\mathbf{q}$ defined in \eqref{eq:total feasible region}. Substituting \eqref{eq:fixed_load_uncertainty_model} into the $\mathbf{b}_\mathbf{q}$ definition yields
\begin{align}
 &\|A_{\mathcal{M}} W\|_2 + A_{\mathcal{M}} \mathbf{g} + A_{\mathcal{N}}^+ \mathbf{P}^+_{\mathcal{N}} + A_{\mathcal{N}}^- \mathbf{P}^-_{\mathcal{N}} \leq \bar{\mathbf b}_{\mathbf q} -H\Delta \boldsymbol{\zeta} \nonumber\\
 &\triangleq~~  g(W,\mathbf g,\mathbf P_{\mathcal N}^\pm)\leq \bar{\mathbf b}_{\mathbf q} -H\Delta \boldsymbol{\zeta},\label{eq:bq_uncertain} 
\end{align}
where ${\bar{\mathbf b}_{\mathbf q}} = \mathbf{c}-H\bar{\mathbf s}^{\mathrm{fixed}}-B \mathbf{q}$. Enforcing~\eqref{eq:bq_uncertain} for all $\boldsymbol{\zeta}\in\mathcal U(\Gamma)$ is equivalent to
\begin{equation}
g_\ell(\cdot)\ \le\ \bar b_{\mathbf q,\ell} - \max_{\boldsymbol{\zeta}\in\mathcal U(\Gamma)} h_\ell^\top \boldsymbol{\zeta},\quad \forall \ell\in\{1,\dots,L\}.
\label{eq:scalar_robust_constraints}
\end{equation}
Here $L$ denotes the number of rows of \eqref{eq:bq_uncertain}. For each row $\ell$, $g_\ell(\cdot)$, $\bar b_{\mathbf q,\ell}$ denote the $\ell$-th components of $\mathbf g$, $\bar{\mathbf b}_{\mathbf q}$, and $h_\ell^\top$ denotes the $\ell$-th row of $[H\Delta]$. 
Since $h_\ell$ is not a variable and $\mathcal U(\Gamma)$ is symmetric and bounded, the max term in \eqref{eq:scalar_robust_constraints} can be computed offline by aligning $\boldsymbol{\zeta}$ with the signs of $h_\ell$. Thus, defining the precomputed worst-case bound
\begin{equation}
\delta_\ell(\Gamma)\ \triangleq\ \max_{\boldsymbol{\zeta}\in\mathcal U(\Gamma)} h_\ell^\top \boldsymbol{\zeta}=
\max_{\substack{s\in\mathbb{R}^{N}\\ 0\le s\le \mathbf 1,\ \mathbf 1^\top s\le \Gamma}}
\sum_{j=1}^{N} |h_{\ell j}|\, s_j.
\label{eq:delta_abs_form}
\end{equation}
The maximum corresponds to selecting up to $\Gamma$ largest components of $|h_{\ell j}|$ at full weight (and, if $\Gamma$ is not an integer, one additional component fractionally). Analytically,
\begin{equation}
\delta_\ell(\Gamma)
=
\sum_{r=1}^{k} \gamma_{\ell,r}
\;+\;
\theta\, \gamma_{\ell,k+1},
\label{eq:delta_sorted}
\end{equation}
where $\gamma_{\ell,1}\ge \gamma_{\ell,2}\ge \cdots \ge \gamma_{\ell,N}$ denote the entries of $\{|h_{\ell j}|\}_{j=1}^{N}$ sorted in non-increasing order, and $k=\lfloor \Gamma\rfloor$, and $\theta =\Gamma - k\in[0,1)$. Applying this to \eqref{eq:bq_uncertain} gives us the robust counterpart of \eqref{eq:main inequality} as 
\begin{subequations}\label{eq:uncertainty inequality}
\begin{align}
&\|A_{\mathcal{M}} W\|_2 + A_{\mathcal{M}} \mathbf{g} + A_{\mathcal{N}}^+ \mathbf{P}^+_{\mathcal{N}} + A_{\mathcal{N}}^- \mathbf{P}^-_{\mathcal{N}} \leq \bar{\mathbf b}_{{\mathbf q}\Gamma} \\
&\bar{\mathbf b}_{{\mathbf q}\Gamma} = \mathbf{c}-H\bar{\mathbf s}^{\mathrm{fixed}}-B \mathbf{q} -\delta(\Gamma)
\end{align}
\end{subequations}
Note that the uncertainty model in \eqref{eq:fixed_load_uncertainty_model}, \eqref{eq:budgeted_box_set} will lead to smaller DOEs as the nominal feasibility constraint is tightened in \eqref{eq:uncertainty inequality}, with higher $\Gamma$ leading to greater tightening.

\subsection{Fairness for DOE}
\label{subsec:fairness_directional}
Predictably, maximizing flexibility volume in Problem \eqref{eq:main surr. prob} can yield DOEs with large disparity, which may be in conflict with customer weights (e.g., relative values of connection capacity, contracted exports and imports, or other policy-determined access rights). To mitigate such disparity, we now present a principled method to regularize the DOE design. 

In the following, we model a certificate-based close-to-fair allocation of maximum permissible flexibility injections among customers, considering export (positive injection) and import (negative injection) separately. We introduce two directional certificate vectors $\mathbf P_{\mathcal M}^{+}, 
\mathbf P_{\mathcal M}^{-}\in\mathbb{R}^{n_{\mathcal M}},$ which represent certified aggregate export and import headroom levels for the coordinated set $\mathcal{M}$, respectively. To ensure feasibility, they must satisfy the DOE feasibility constraint \eqref{eq:main inequality} when combined with maximum injections by the non-coordinated customers. These certificate vectors are therefore constrained by the final polytopic DOE through \eqref{eq:env_feas_plus}--\eqref{eq:env_feas_minus}, while their aggregate import/export levels are linked to the ellipsoidal support values through \eqref{eq:sum_dom1}--\eqref{eq:sum_dom2}. Thus, the fairness quantities used below are witnessed by feasible operating points of the published coordinated DOE. This is formalized by the following constraints:

\begin{subequations}
\label{eq:directional}
\begin{align}
&A_{\mathcal{M}} \mathbf P_{\mathcal M}^{+}
+ A_{\mathcal{N}}^+ \mathbf{P}^+_{\mathcal{N}} + A_{\mathcal{N}}^- \mathbf{P}^-_{\mathcal{N}}
\leq \mathbf{b}_\mathbf{q},
\label{eq:env_feas_plus}\\
&A_{\mathcal{M}} \mathbf P_{\mathcal M}^{-}
+ A_{\mathcal{N}}^+ \mathbf{P}^+_{\mathcal{N}} + A_{\mathcal{N}}^- \mathbf{P}^-_{\mathcal{N}}
\leq \mathbf{b}_\mathbf{q} \label{eq:env_feas_minus}\\
&\mathbf 1^\top \mathbf P_{\mathcal M}^{+} 
\ge \mathbf 1^\top \mathbf g + \|W^\top \mathbf 1\|_2,\label{eq:sum_dom1} \\
&\mathbf 1^\top \mathbf P_{\mathcal M}^{-} 
\le \mathbf 1^\top \mathbf g - \|W^\top \mathbf 1\|_2,
\label{eq:sum_dom2}
\end{align}
\end{subequations}
Constraints \eqref{eq:sum_dom1},\eqref{eq:sum_dom2} ensure that $(\mathbf P_{\mathcal M}^{+},\mathbf P_{\mathcal M}^{-})$ provide aggregate export/import levels that are at least as permissive as those achievable by the coordinated ellipsoid $(W,\mathbf g)$ along the aggregated direction $\mathbf 1$.\footnote{For ellipsoid $\{Wu+\mathbf g:\|u\|_2\le 1\}$, its support function along the direction $d$ gives:
$
\max_{\|u\|_2\le 1}$ $d^\top(Wu+\mathbf g)=$ $d^\top \mathbf g+\|W^\top d\|_2$,
$\min_{\|u\|_2\le 1}$ $d^\top(Wu+\mathbf g)=$ $d^\top \mathbf g-\|W^\top d\|_2$.
}

The total permissible export and import flexibility can be denoted as
\begin{equation}
F^{\pm} \;=\; \pm\mathbf 1^\top \mathbf P_{\mathcal M}^{\pm} \;\pm\; \sum_{i\in\mathcal N} P_i^{\pm}.
\label{eq:Fpm_defs}
\end{equation}

Let $\omega_k^{+},\omega_k^{-}\ge 0$ denote pre-defined export and import weights for each participant $k\in \{\mathcal M\}\cup\mathcal N$, that signify their importance when considered within a fairness paradigm. For set $\mathcal M$, $\omega_{\mathcal M}^{\pm}$ is the group-level weight. Define normalized weights
\begin{equation}
\alpha_k^{\pm}
=\frac{\omega_k^{\pm}}{\omega_{\mathcal M}^{\pm}+\sum_{j\in\mathcal N}\omega_j^{\pm}},
\qquad k\in\{\mathcal M\}\cup\mathcal N .
\label{eq:alpha_dir}
\end{equation}

For user-chosen fairness parameters $\sigma^{+},\sigma^{-}\in[0,1]$, we impose the following export and import fairness constraints
\begin{subequations}
 \begin{align}
&\pm P_i^{\pm} \ge (1-\sigma^{\pm})\,\alpha_i^{\pm}\,F^{\pm},
\quad \forall i\in\mathcal N, \\
&\pm \mathbf 1^\top \mathbf P_{\mathcal M}^{\pm}
\ge (1-\sigma^{\pm})\,\alpha_{\mathcal M}^{\pm}\,F^{\pm},
\end{align} 
\label{eq:fair_export_import}
\end{subequations}

These constraints require each non-coordinated customer and the coordinated cohort to receive at least a $(1-\sigma^\pm)$ fraction of their weighted share of the certified total export/import headroom. At $\sigma^\pm=1$, \eqref{eq:fair_export_import} is trivially satisfied, and no fairness requirement is imposed, whereas $\sigma^\pm=0$, the strictest weighted-share requirement is enforced within the proposed certificate-based fairness model.

Note that our fairness model is spatial and snapshot-based, reflecting a single operating interval. It is related to other dispersion-based fairness models such as $\epsilon$-fairness and Jain index \cite{sundar2025parametric, jain1984quantitative} as reported in \cite{bhadoriya2025family,rubbers2025fairness}. It is worth mentioning that fairness is a multifaceted concept, and richer notions (e.g., temporal fairness) are beyond the scope of the present paper. We defer such extensions to future work.

\textbf{DOE Design with Robustness and Fairness:}
We now present an extension of \eqref{eq:main surr. prob}, where maximal volume DOE construction additionally encodes robustness against budgeted load uncertainty (Section \ref{subsec:robust_fixed_load}), and fairness constraint (Section \ref{subsec:fairness_directional}). The modified formulation is given below:

\begin{subequations}
\label{eq:final_problem}
\begin{align}
 \max_{\substack{
P^\pm_i, \mathbf{q},W, \mathbf{c},
\mathbf{P}^\pm_\mathcal{M} 
}}
\quad 
 & \log \det(W) + \sum_{i \in \mathcal{N}} \log (P_i^+ - P_i^-) \\[6pt]
 \text{s.t.} \quad 
 & \eqref{eq:zero_non}, \eqref{eq:zero_coord}, \eqref{eq:uncertainty inequality}, \eqref{eq:directional}, \eqref{eq:Fpm_defs}, \eqref{eq:fair_export_import} \\ 
 & W \succeq 0.
\end{align}
\end{subequations}
Here \eqref{eq:uncertainty inequality} models the load uncertainty and \eqref{eq:directional}, \eqref{eq:Fpm_defs}, \eqref{eq:fair_export_import} models the fairness constraint. Making $\Gamma =0$ (no uncertainty) and $\sigma^\pm=1$ reduces Problem \eqref{eq:final_problem} to the original formulation \eqref{eq:main surr. prob}. 
Once the DOE design problem in \eqref{eq:final_problem} is solved, the final DOE for the coordinated set is computed as explained in Section~\ref{sec:poly_from_ellipsoid}.

\section{Numerical Experiments and Validation}
\label{sec:case}
\begin{figure}
 \centering
 \includegraphics[width=0.75\linewidth]{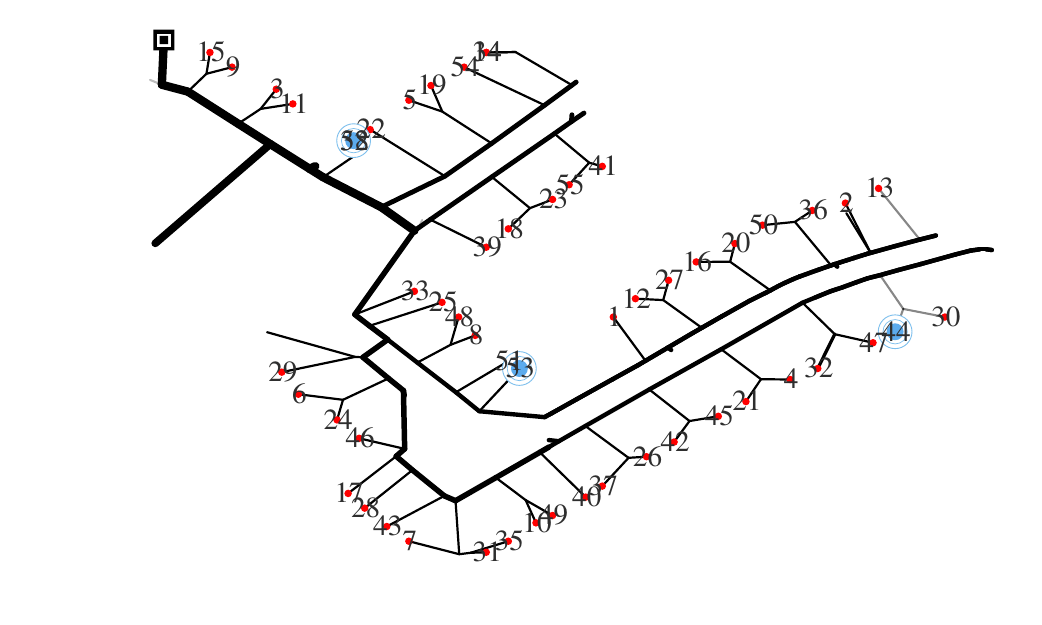}
 \caption{European LV feeder with highlighted coordinated loads.}
 \label{fig:feeder}
\end{figure}

To validate our proposed DOE framework, we first conduct numerical experiments on the European Low Voltage Test Feeder~\cite{ieee2015european}, a benchmark system widely used for distributed energy resource integration and flexibility studies. The network topology is shown in Fig.~\ref{fig:feeder}.

The numerical experiments were performed on a 13th Gen Intel(R) Core(TM) i9-13950HX CPU at 2.20GHz with 64GB of RAM. For the non-convex AC optimal power flow tests, we used the interior-point solver \texttt{IPOPT v3.14}, while the proposed convex formulation was solved using \texttt{MOSEK v10.2}. 

\subsection{Base Case Description}
\label{sec:basecase}
The base case is configured to reflect DOE roles in safe operating conditions of the LV feeder. For better visualization, in this case, we assume there are only three coordinated loads, shown in blue circles in Fig. \ref{fig:feeder}, that are equipped for coordinated operation, the rest being uncoordinated. All customers are assumed to be homogeneously having $\pm5~\text{kW}$ maximum flexible active power and reactive set point that can be set between $\pm2~\text{kVAR}$. Additionally, each customer has a random fixed consumption up to $1\text{kW}$ at 0.95 power factor. The statutory voltage range is kept at $1\pm(5\%)$ per unit. 
In all numerical experiments, the line-flow circle in \eqref{eq:line thermal circle} is represented by an octagonal inner approximation, corresponding to $2\rho=8$ facets in \eqref{eq:line flow approximation}. In the initial experiments, fairness and load uncertainty are not considered, and DOEs are designed using Problem \eqref{eq:main surr. prob}.

\begin{figure}
 \centering
 \includegraphics[width=0.72\linewidth]{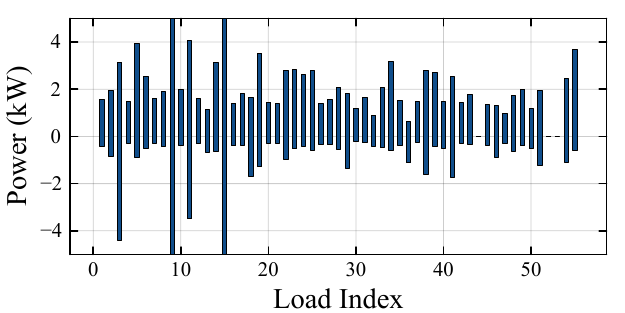}
 \caption{DOEs for non-coordinated loads. The DOE for coordinated customers \#44, \#52, and \#53 is not shown here.}
 \label{fig:DOE non}
\end{figure}

\begin{figure}[ht!]
 \centering
 \includegraphics[clip, trim=0cm 0cm 0cm 1cm, width=0.80\linewidth]{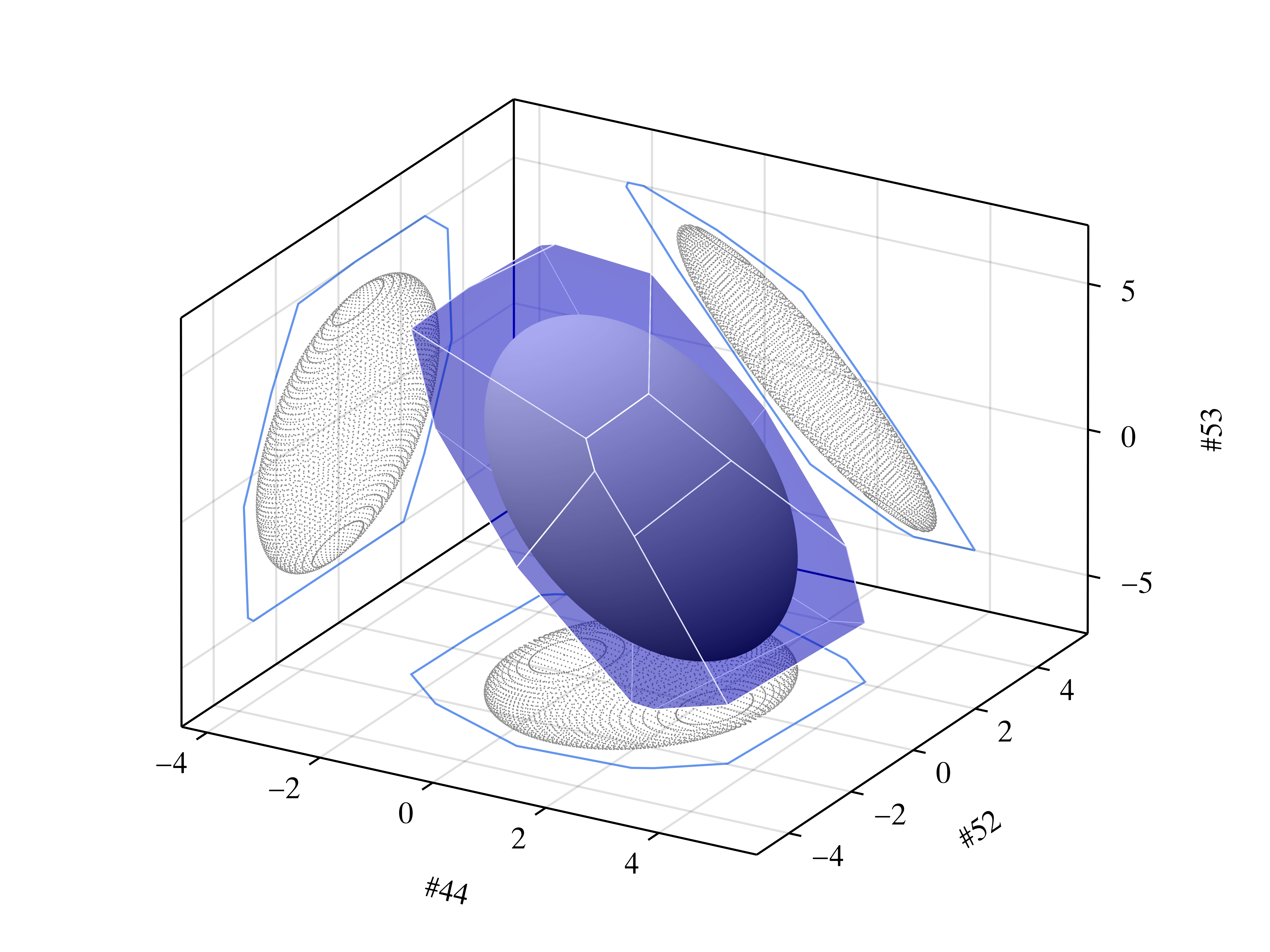}
 \caption{Coordinated polytopal DOE for customers \#44, \#52, and \#53, inscribed ellipsoid used during DOE design, and their projections on the three pairwise coordinate planes for better visualization. Axes represent kW.}
 \label{fig:DOE comm}
\end{figure}

The DOEs for non-coordinated and coordinated load types are shown in Fig.~\ref{fig:DOE non} and Fig. \ref{fig:DOE comm}, respectively. Note that DOEs for non-coordinated customers include intervals for flexible active power, while the coordinated cohort's DOE is a 3-D polytope obtained via the construction in Section~\ref{sec:poly_from_ellipsoid}.

\subsection{Stress test for feasibility}
\label{sec:adversarial test}
To validate that DOEs computed using the linearized power flow model (see \eqref{eq:volt_eq}, \eqref{eq:flow_eq}) are compatible with the \emph{nonlinear} AC power flow equations, we perform an adversarial feasibility analysis. For the DOEs (Fig.~\ref{fig:DOE non}--\ref{fig:DOE comm}), we solve a sequence of AC power flow optimization problems in which worst-case injections within the designed DOE limits are selected to drive extreme operating conditions, i.e., to maximize or minimize the voltage magnitude at each node, and to maximize the apparent power flow at each line. 

\begin{figure}
 \centering
 \includegraphics[width=0.75\linewidth]{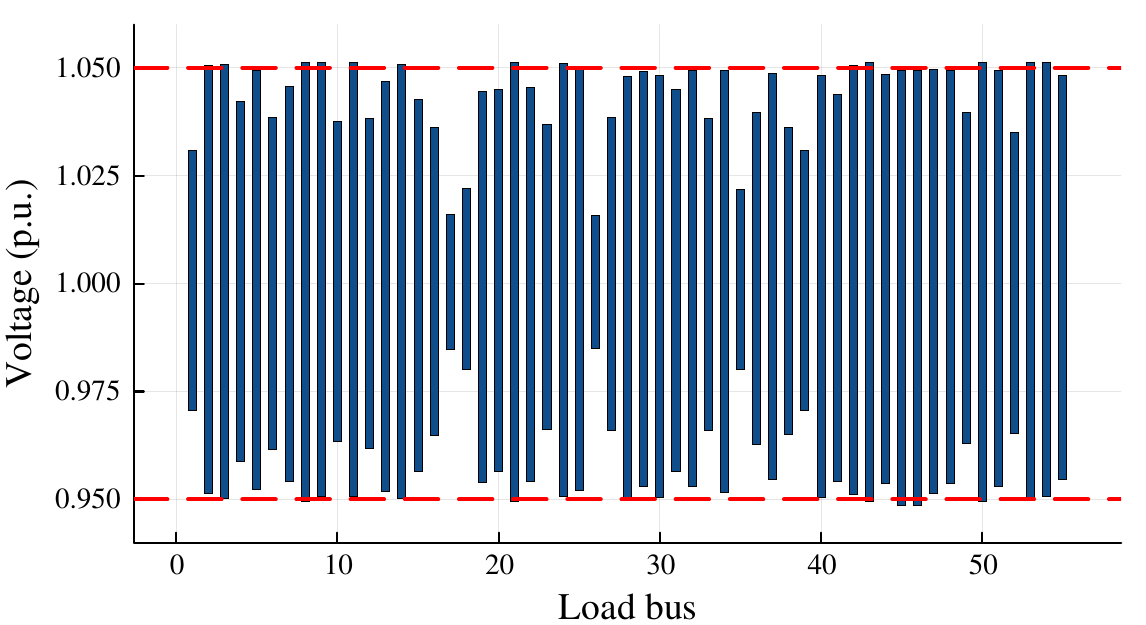}
 \caption{Adversarial AC power flow voltage magnitudes (minimum and maximum) attained over DOE-admissible injections }
 \label{fig:voltage level envelope}
\end{figure}

\begin{figure}
 \centering
 \includegraphics[width=0.75\linewidth]{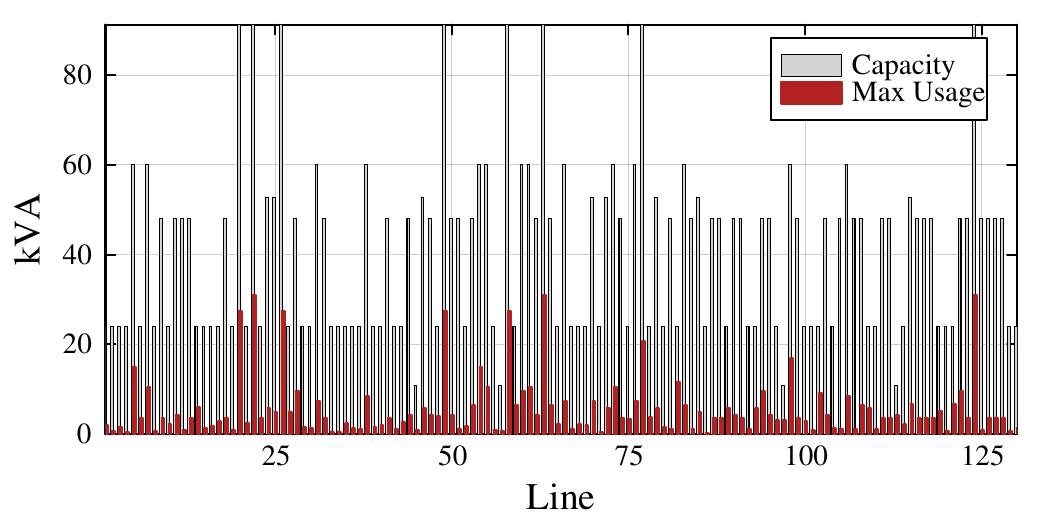}
 \caption{Adversarial maximum apparent line flow under AC power-flow over DOE-admissible injections (red bars). Grey bars show line ratings.}
 \label{fig:line cap utilization}
\end{figure}

Importantly, in these adversarial problems, we do not enforce voltage and thermal limits as constraints. Fig.~\ref{fig:voltage level envelope} reports, for each node, the minimum and maximum voltage magnitude attained over the adversarial optimizations. Fig.~\ref{fig:line cap utilization} reports, for each line, the maximum apparent-power loading attained and the thermal rating. No thermal-limit violation is observed, and the maximum over-voltage and under-voltage violations are \(0.0012\) p.u. and \(0.0013\) p.u.,
respectively. This indicates that over the DOE-admissible injections, voltages and line loadings are consistent with the network operational limits.

\subsection{Impact of Load Coordination}
\label{sec:load coordination}
To quantify the impact of load coordination, we systematically vary the share of coordinated customers, letting their number $n_\mathcal{M}$ range from a single customer up to the entire population. For each case, the remaining customers act as non-coordinated, and we run 10 randomized trials to solve our DOE design problem under different grouping combinations. 

\begin{figure}[ht!]
 \begin{subfigure}[b]{0.49\linewidth}
 \includegraphics[width=\linewidth]{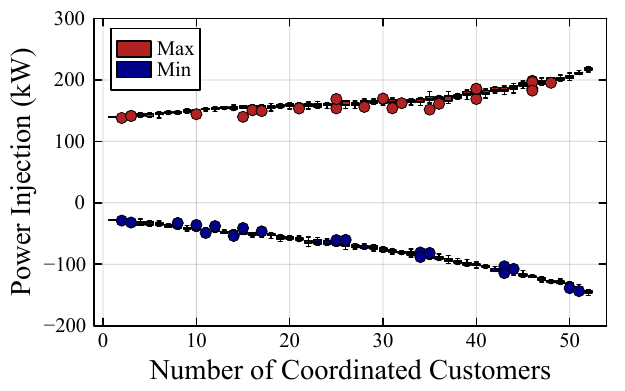}
 \caption{}
 \label{fig:power ranges}
 \end{subfigure}
 \hfill 
 \begin{subfigure}[b]{0.49\linewidth}
 \includegraphics[width=\linewidth]{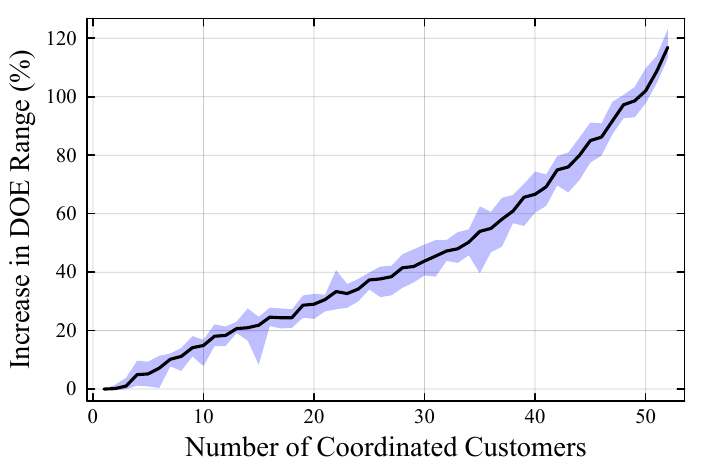}
 \caption{}
 \label{fig:power ranges increase}
 \end{subfigure}
 \caption{Coordination evaluation as the increase in aggregate power ranges, a) Max and min aggregate power injections within DOEs, b) Percentage increase in aggregate power ranges w.r.t. no coordination case. In panel (b), the solid black curve reports the average over \(10\) random coordinated-customer groupings, while the blue shaded band shows the min--max range across those groupings.
}
 \label{fig:DOE vs. coordination}
\end{figure}

To obtain a kW-based metric for the total DOE , we post-processed each DOE by solving two linear programs that maximize and minimize the aggregate active-power injection $\sum_j P_j$, subject to the DOE polyhedral constraints for coordinated customers and the corresponding box/range constraints for non-coordinated customers. In this post-processing step, the coordinated DOE is represented by the residual polytope constructed in Section~\ref{sec:poly_from_ellipsoid}. The resulting injection ranges (Fig.~\ref{fig:power ranges}) show that coordination substantially expands the aggregate operating span. To facilitate comparison across cases, Fig.~\ref{fig:power ranges increase} reports the percentage increase of the aggregate injection range relative to the non-coordinated baseline. In Fig.~\ref{fig:power ranges increase}, the solid black curve reports the average aggregate DOE expansion over \(10\) random coordinated-customer groupings for each coordination share, while the blue shaded band reports the min--max range across these groupings. The results indicate that even partial coordination can deliver meaningful gains; for example, coordinating 30\% of customers (16 customers in our study) increases the achievable aggregate injection range by approximately 25\% on average across the randomized trials.

\begin{figure}[ht!]
 \centering
 \includegraphics[width=0.65\linewidth]{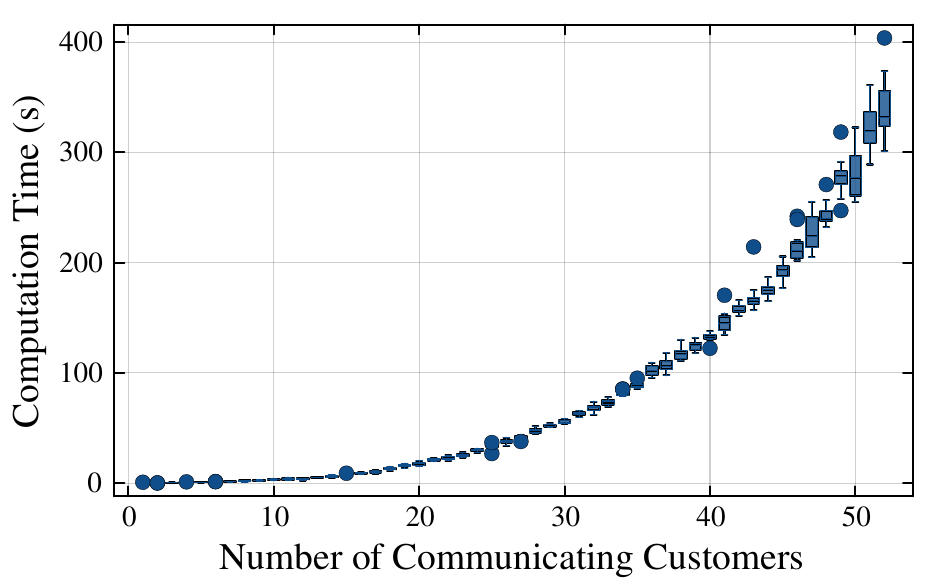}
 \caption{Optimization problem computation time vs. the number of coordinated loads. Each bar shows the range of computation times across 10 randomized customer groupings for each coordination level.}
 \label{fig:comput time}
\end{figure}

The computation time in Fig.~\ref{fig:comput time} shows an increase with the number of coordinated customers. This trend is expected due to the growth in problem size, particularly the PSD matrix variable and the associated log-determinant constraint in \eqref{eq:main surr. prob}. 

Next, we show simulation results for the DOE design under uncertainty in fixed loads, and with fairness across customers.

\subsection{Uncertainty in Fixed Loads}
\label{sec:case_fixed_load_uncertainty}

This case study evaluates the impact of fixed-load forecast uncertainty on the DOEs, under the robust formulation in Section \ref{subsec:robust_fixed_load}. 

We generate a baseline fixed-load profile by randomly selecting each customer's nominal active and reactive powers independently from uniform bounded ranges: $\bar p_i^{\mathrm{fixed}} \sim \mathcal{U}(-2.5,\,2.5)$ kW and $\bar q_i^{\mathrm{fixed}} \sim \mathcal{U}(-1,\,1)$ kVAr. To emulate different operating conditions, we scale all nominal fixed loads ($p^{\mathrm{fixed}}, q^{\mathrm{fixed}}$) by a common loading factor $\{0.5,\,1.0,\,2.0\}$. This yields three loading regimes: low, moderate, and high. Fixed-load uncertainties are modeled as bounded deviations around the nominal values, with bounds chosen proportionally to the nominal loads, $\Delta^p_i = \eta \, \bigl|\bar p^{\mathrm{fixed}}_i\bigr|,
$ and $ \Delta^q_i = \eta \, \bigl|\bar q^{\mathrm{fixed}}_i\bigr|$. We consider uncertainty magnitudes $\eta \in \{0.1, 0.2, 0.3\}$, and a Bertsimas--Sim budget parameter $\Gamma \in \{0,5,10,15,20\}$ in \eqref{eq:budgeted_box_set}. The flexibility limit of each customer is randomly sampled from the set $\{0,3,5,7\}\,\mathrm{kW}$ to emulate heterogeneity. The selected values of \(\eta\) and \(\Gamma\) are used to trace the robustness--performance trade-off. In an operational deployment, these parameters can be replaced by values informed by historical forecast-error data. Increasing either \(\eta\) or \(\Gamma\) enlarges the uncertainty set and reserves more network headroom for robustness.

\begin{figure}[ht!]
 \centering
 \includegraphics[width=1.05\linewidth]{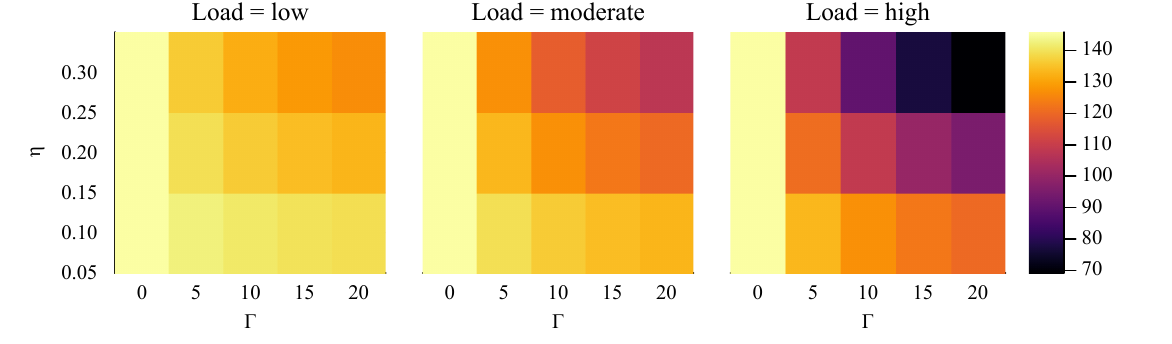}
 \caption{Aggregate network-level DOE range (kW) under fixed-load uncertainty. The uncertainty magnitude $\eta$ scales the component-wise deviation bounds $\Delta$ in \eqref{eq:fixed_load_uncertainty_model}, while the budget parameter $\Gamma$ in \eqref{eq:budgeted_box_set} controls conservatism by limiting how many components deviate simultaneously.}
 \label{fig:uncertainty_heatmap}
\end{figure}

To summarize flexibility in engineering units, we report the aggregate network-level DOE range (in kW), for different scenarios of $\eta$ and $\Gamma$ and over the three fixed-load levels, in Figure~\ref{fig:uncertainty_heatmap}.  As expected, increasing either $\eta$ or $\Gamma$ reduces the aggregate DOE range. Moreover, the relative impact of uncertainty is amplified under higher loading. Specifically, high uncertainty leads to a reduction in the aggregate DOE range by about 13\% in the low-loading regime, while the reduction is nearly 53\% in the high-loading regime.

\begin{figure*}[bt!]
 \centering
 \begin{subfigure}{0.66\columnwidth}
  \centering
  \includegraphics[width=\linewidth]{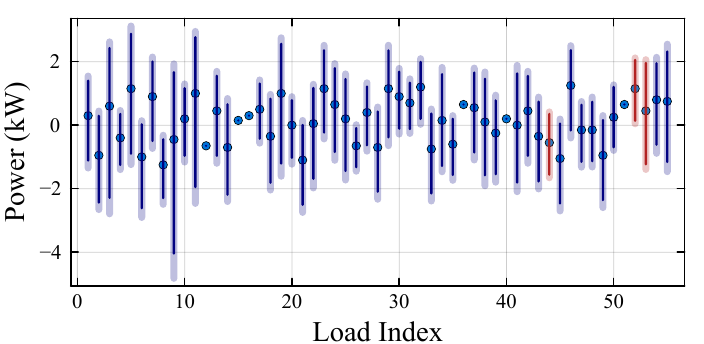}
  \caption{Low loading.}
  \label{fig:doe_low}
 \end{subfigure}
 \begin{subfigure}{0.66\columnwidth}
  \centering
  \includegraphics[width=\linewidth]{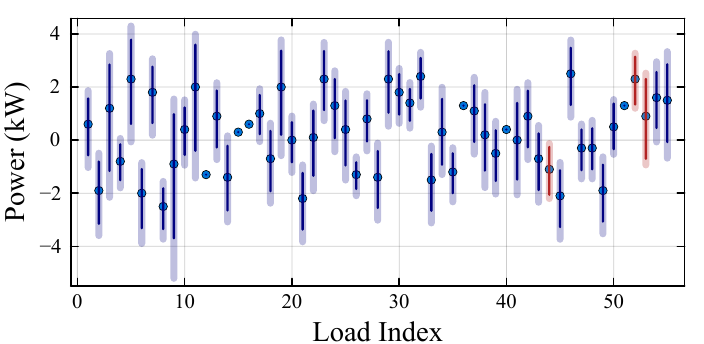}
  \caption{Moderate loading.}
  \label{fig:doe_moderate}
 \end{subfigure}
 \begin{subfigure}{0.66\columnwidth}
  \centering
  \includegraphics[width=\linewidth]{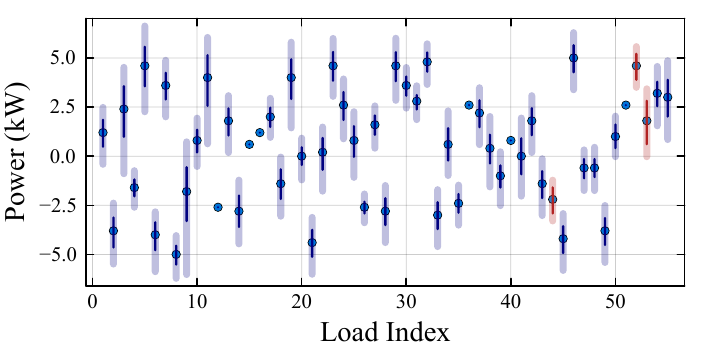}
  \caption{High loading.}
  \label{fig:doe_high}
 \end{subfigure}
 \caption{Customer-level DOE intervals around the base load for three loading regimes. In each panel, we overlay the least conservative uncertainty setting $(\eta_{\min},\Gamma_{\min})$ (lighter) and the most conservative setting $(\eta_{\max},\Gamma_{\max})$ (darker). Blue: non-coordinated DOEs; red: weight-proportional visualization of the coordinated group's aggregate DOE.}
 \label{fig:doe_three_levels}
\end{figure*}

To provide intuition at the customer level, Figures~\ref{fig:doe_low}--\ref{fig:doe_high} overlay two representative DOE realizations for each loading regime: (i) low uncertainty $(\eta_{\min}, \Gamma_{\min})$ and (ii) high uncertainty $(\eta_{\max}, \Gamma_{\max})$. For each customer, we plot the base load and the DOE interval as flexibility around the base load. The least conservative case is shown with lighter styling, while the most conservative case is shown with darker styling. These plots illustrate that robustification to uncertainty reduces DOE across customers, with the reduction being more pronounced at higher loading levels.

\subsection{Impact of fairness}
\label{case:fairness}
In this section, we study how the proposed directional fairness constraints~\eqref{eq:fair_export_import} affect the distribution of export/import flexibility across customers and the overall size of the resulting envelopes. In this study, the selection of coordinated loads is similar to Section \ref{sec:basecase}. 
Customer weights $\boldsymbol{\omega}^{\pm}$ are interpreted as nominal export/import access rights (here taken proportional to connection ratings). In this case study, weights are selected as in Section \ref{sec:case_fixed_load_uncertainty}, i.e., sampled from $\{0,3,5,7\}$ to emulate heterogeneity, and the fairness parameters are set symmetrically, $\sigma^+=\sigma^-=\sigma$, with $\sigma$ varied in steps of $0.1$. Note that a higher $\sigma$ implies a lower fairness constraint. For our simulations, we track two metrics:
\begin{itemize}[leftmargin=0pt,itemindent=*,align=left]
 \item \textit{Average envelope size:} The geometric mean of the joint envelope size, computed as $\left(\mathrm{Vol}(
 \mathcal{P}_{\{\mathbf{p}^{\mathcal{M}}\}})\cdot \mathrm{Vol}(\mathcal{H}_{\{\mathbf{p}^{\mathcal{N}}\}})\right)^{1/n_{\mathrm{act}}}$, where $\mathcal{P}_{\{\mathbf{p}^{\mathcal{M}}\}}$ denotes the coordinated customers' polytopal DOE, $ \mathcal{H}_{\{\mathbf{p}^{\mathcal{N}}\}}$ shows the hyperrectangular DOE for non-coordinated customers, and $n_{\mathrm{act}}$ is the number of participants with nonzero weight. This normalization yields a kW-scale quantity that is comparable across cases. We computed the volume numerically using the \texttt{VolEsti} library~\cite{volesti}, which estimates volumes via randomized sampling–based algorithms.
 \item \textit{Fairness index:} We measure fairness over the weight-normalized DOE allocations. First, we compute the allocated maximum export and import flexibility as
 \[
 a_k^+ = \begin{cases}
 \mathbf{1}^\top \mathbf{P}^+_{\mathcal M}, & k=\mathcal M,\\
 P_k^+, & k\in\mathcal N,
 \end{cases}
 \qquad
 a_k^- = \begin{cases}
 -\mathbf{1}^\top \mathbf{P}^-_{\mathcal M}, & k=\mathcal M,\\
 -P_k^-, & k\in\mathcal N,
 \end{cases}
 \]
 and consider the weights $\alpha_k^\pm$, defined in \eqref{eq:alpha_dir}. Excluding participants with $\alpha_k^+ + \alpha_k^- = 0$, we determine the weight-normalized DOE allocation as:
 \[
 x_k = \frac{a_k^+ + a_k^-}{\alpha_k^+ + \alpha_k^-},
 \qquad k\in\{\mathcal M\}\cup\mathcal N.
 \] Note that $x_k$ measures the total injection range $(a_k^+ + a_k^-)$ per unit of weight $(\alpha_k^+ + \alpha_k^-)$. We measure fairness using the Gini coefficient \footnote{For a vector $x\in\mathbb{R}^n_{+}$ with mean $\bar{x}$, the Gini coefficient is
$G=\frac{\sum_{i=1}^{n}\sum_{j=1}^{n}|x_i-x_j|}{2n^2\bar{x}}$.} \cite{gini1936measure} for $x_k$s. 
 A smaller Gini value indicates a more equitable allocation.
\end{itemize}

\begin{figure}[ht!]
 \begin{subfigure}[b]{0.49\columnwidth}
 \includegraphics[width=\linewidth]{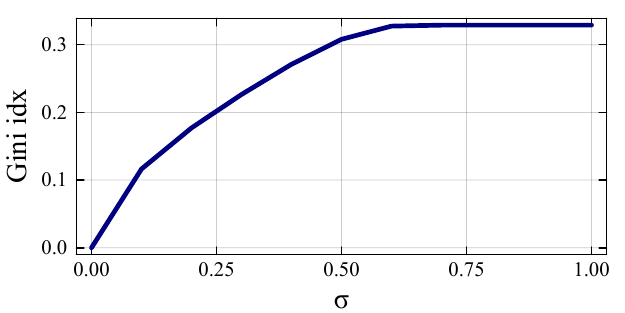}
 \caption{}
 \label{fig:gini_vs_sigma}
 \end{subfigure}
 \hfill 
 \begin{subfigure}[b]{0.49\columnwidth}
 \includegraphics[width=\linewidth]{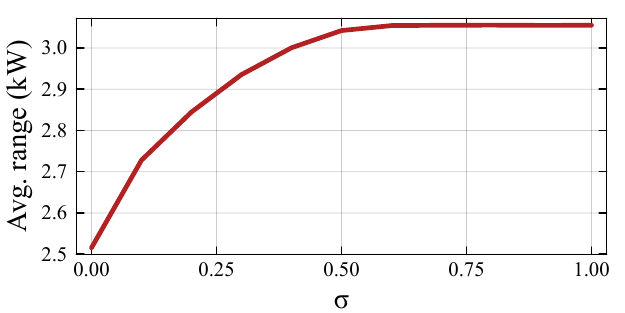}
 \caption{}
 \label{fig:size_vs_sigma}
 \end{subfigure}
 \caption{Effect of fairness-relaxation parameter $\sigma^\pm =\sigma$ on (a) disparity relative to weights and (b) the normalized envelope size. Smaller \(\sigma\) corresponds to stronger fairness enforcement, while larger \(\sigma\) relaxes the fairness constraint.}
 \label{fig:fairness_study}
\end{figure}

Fig.~\ref{fig:fairness_study} summarizes the results. As $\sigma$ decreases (tighter requirement in constraint \eqref{eq:fair_export_import}), the Gini index decreases, indicating reduced disparity relative to weights, while the envelope-size decreases, reflecting the expected trade-off between fairness and feasible headroom. In particular, enforcing the strictest fairness setting results in an approximately $14\%$ reduction in the average DOE range relative to the $\sigma=0.25$ case.

\begin{figure*}[tb!]
 \begin{subfigure}{0.68\columnwidth}
 \includegraphics[width=\linewidth]{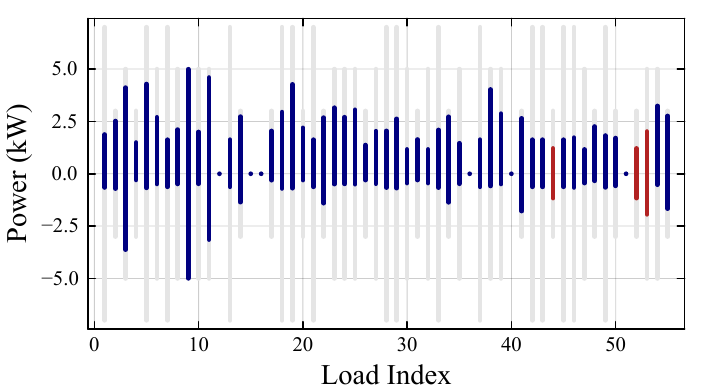}
 \caption{$\sigma^\pm=0.5$}
 \end{subfigure}
 \begin{subfigure}{0.68\columnwidth}
 \includegraphics[width=\linewidth]{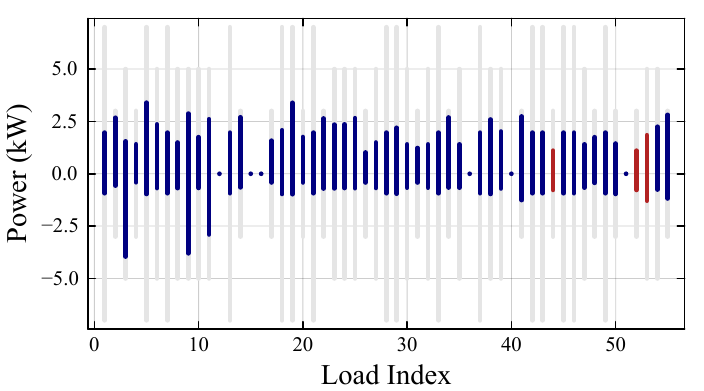}
 \caption{$\sigma^\pm=0.3$}
 \end{subfigure}
 \begin{subfigure}{0.68\columnwidth}
 \includegraphics[width=\linewidth]{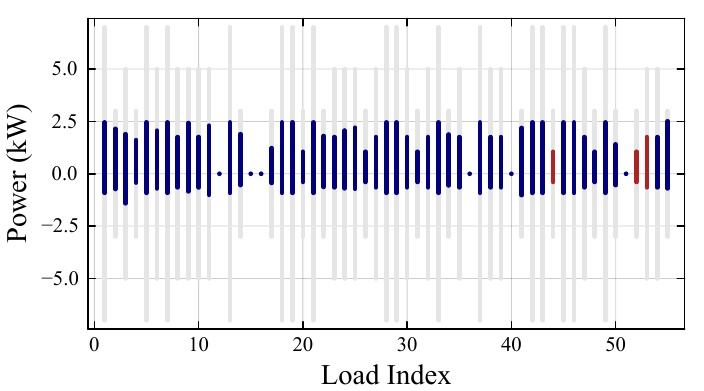}
 \caption{$\sigma^\pm=0.1$}
 \end{subfigure}
 \caption{Illustration of the effect of fairness constraints on customer-level allocations. Gray: weights; blue: non-coordinated DOEs; red: weight-proportional visualization of the coordinated group's aggregate DOE.}
  \label{fig:fairness_DOE_impact_study}
\end{figure*}

To illustrate how fairness reshapes the allocations, Fig.~\ref{fig:fairness_DOE_impact_study} plots customer-level intervals for several $\sigma$ values. Gray bars show weight limits, and the blue bars are the actual DOE intervals assigned to non-coordinated customers. Coordinated customers instead receive a coupled polytopal DOE; therefore, the red bars provide only a weight-proportional visualization of the aggregate certificate headroom. The relationship between these certificate quantities and the ranges induced by the final polytope is examined in Section~\ref{sec:benchmark_pq}. As expected, smaller $\sigma$ (high fairness) values raise the minimum guaranteed headroom of participants and compress the spread.

While the results presented till here included active power DOEs, we next present results for  active--reactive DOEs discussed in Section \ref{sec:pq_extension}.Furthermore we benchmark our results against the \(P\)--\(Q\) DOE method in \cite{mahmoodi2023capacity}, over a large three-phase feeder.

\subsection{Active--Reactive DOE, Benchmarking and Scalability}
\label{sec:benchmark_pq}

We select \cite{mahmoodi2023capacity} as the benchmark because it constructs customer-level active--reactive DOE polygons and is therefore directly comparable with the \(P\)--\(Q\) extension in Section \ref{sec:pq_extension}.

The comparison is carried out on a larger unbalanced three-phase feeder, network 9 in \cite{espinosa2015dissemination} with \(3589\) buses, \(3588\) lines, and \(124\) load points. Both methods use the same linearized network constraints, the same flexible-customer set, and the same customer capability limits. The customer flexibility limits and fairness weights are selected consistently with the fairness-analysis case in Section~\ref{case:fairness}.

Four cases of the proposed method are considered. Case P1 uses no coordinated
customers and enforces the strictest fairness setting, \(\sigma=0.0\). Case P2 also uses no coordinated customers, but with the fairness setting, \(\sigma=0.35\). This value obtains allocation disparity comparable to the benchmark's ``close-to-equal'' allocation and hence allows flexibility gains to be fairly compared. Cases P3 and P4 use \(\sigma=0.35\) for our proposed framework with \(10\) and \(20\) coordinated customers, respectively and show the additional benefit obtained from coordination.

\begin{figure}[ht]
\centering
\includegraphics[width=0.99\linewidth]{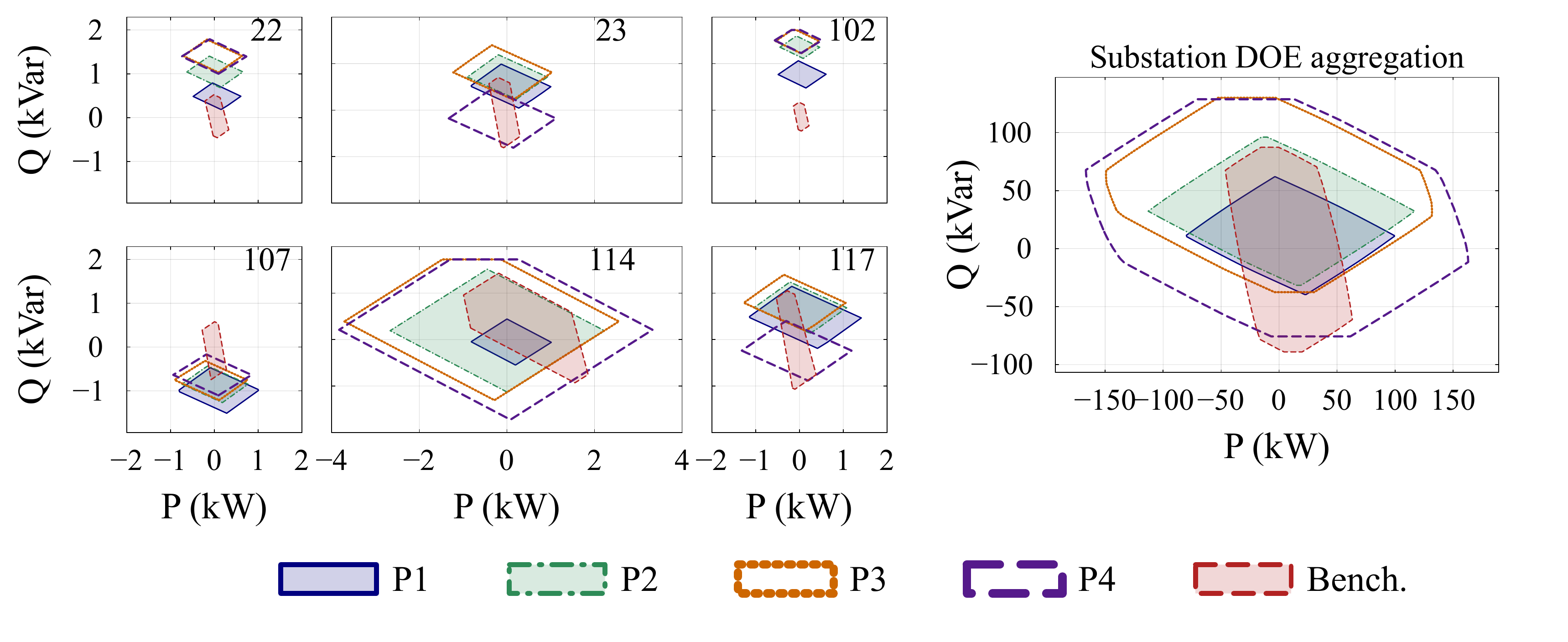}
\caption{Selected customer-level and aggregate \(P\)--\(Q\) DOEs for different settings of the proposed method and the benchmark in~\cite{mahmoodi2023capacity}. Cases P1--P4 denote \((|\mathcal M|,\sigma)=(0,0.00),(0,0.35),(10,0.35),(20,0.35)\), respectively.}
\label{fig:benchmark_A_pq}
\end{figure}
\begin{table}[ht]
\centering
\caption{Benchmark comparison of \(P\)--\(Q\) DOE allocations for different settings of the proposed method. Cases P1--P4 denote
\((|\mathcal M|,\sigma)=(0,0.00),(0,0.35),(10,0.35),(20,0.35)\),
respectively.}
\label{tab:benchmark_A}
\footnotesize

\begin{tabular}{lrrrrr}
\toprule
\textbf{Metric} &
\textbf{P1} &
\textbf{P2} &
\textbf{P3} &
\textbf{P4} &
\textbf{Bench.} \\
\midrule
\(P\) range (kW)
& 179.5 & 229.9 & 281.1 & 329.7 & 109.5 \\
Total area (kVA\(^2\))
& 9531.9 & 15413.0 & 31093.2 & 48169.4 & 13545.8 \\
\(\mathrm{Gini}_{p}\)
& 0.00 & 0.28 & 0.29 & 0.31 & 0.38 \\
\(\mathrm{Gini}_{A}\)
& 0.01 & 0.56 & 0.57 & 0.60 & 0.62 \\
Run time (s)
& 0.5& 0.6& 2.4 & 13.7 & -- \\
Max. OV (p.u.)
& 0.0020 & 0.0027 & 0.0037 & 0.0047 & 0.0000 \\
Max. UV (p.u.)
& 0.0018 & 0.0027 & 0.0037 & 0.0049 & 0.0000 \\
\bottomrule
\end{tabular}

\end{table}

Fig.~\ref{fig:benchmark_A_pq} compares the resulting \(P\)--\(Q\) DOEs. The left panels show the DOE polygons of six randomly selected non-coordinated customers, while the right panels show the corresponding substation-level aggregate \(P\)--\(Q\) regions for the four proposed cases and the benchmark. To complement the DOE plots, Table~\ref{tab:benchmark_A} reports the aggregate active-power range, the aggregate \(P\)--\(Q\) DOE area, the normalized active-range and area range Gini indices, and solution time. The Gini indices are computed after normalizing each customer's allocation by its capability entitlement. Here, the fairness constraints are imposed on active-power import/export headroom, while both $\mathrm{Gini}_\textit{p}$ and $\mathrm{Gini}_{A}$ are calculated ex post to quantify the resulting disparity in active-power ranges and (P)--(Q) DOE areas, respectively.

The results show that our proposed method provides larger active-power operating ranges than the benchmark in all tested cases. P1 increases the total active-power range from \(109.5\) in the benchmark to \(179.5\), while achieving nearly equitable allocation. P2 provides a more directly comparable case: it gives Gini indices close to those of the benchmark, but increases the total active-power range to \(229.9\), which is more than twice the benchmark value. In addition, P2 achieves a larger aggregate \(P\)--\(Q\) area than the benchmark, \(15413.0\) compared with \(13545.8\). The benefit of coordination is observed in P3 and P4, where the aggregate active-power range increases substantially compared with the individual-DOE case P2. With \(20\) coordinated customers, the proposed method provides more than three times the aggregate active-power range of the benchmark, while retaining a comparable level of allocation disparity. The runtime also remains modest: the largest coordinated case is solved in \(13.7\) s.
The observed runtimes indicate that feeder size primarily increases the number of linear network constraints, whereas the coordinated-DOE dimension depends on the cohort size. Thus, the formulation remains suitable for minute-level updates on larger feeders with moderate cohorts; for very large cohorts, simpler inner-set geometries can reduce computation and then be expanded to the final polytope as in \eqref{eq:final polytope}--\eqref{eq:residual rhs}.

We further repeat the worst-case adversarial voltage violation test in Section~\ref{sec:adversarial test} for the benchmarked \(P\)--\(Q\) DOE cases. The last two rows of Table~\ref{tab:benchmark_A} report  small voltage deviations beyond the \(0.95\)--\(1.05\) p.u. voltage band, with the largest over-voltage and under-voltage violations remaining below \(0.005\) p.u. Note that these small boundary violations may not be revealed by standard (average) Monte Carlo sampling. The benchmark has no voltage violation but this is achieved with a much smaller active-power range.

\begin{figure}[t]
\centering
\includegraphics[width=0.75\linewidth]{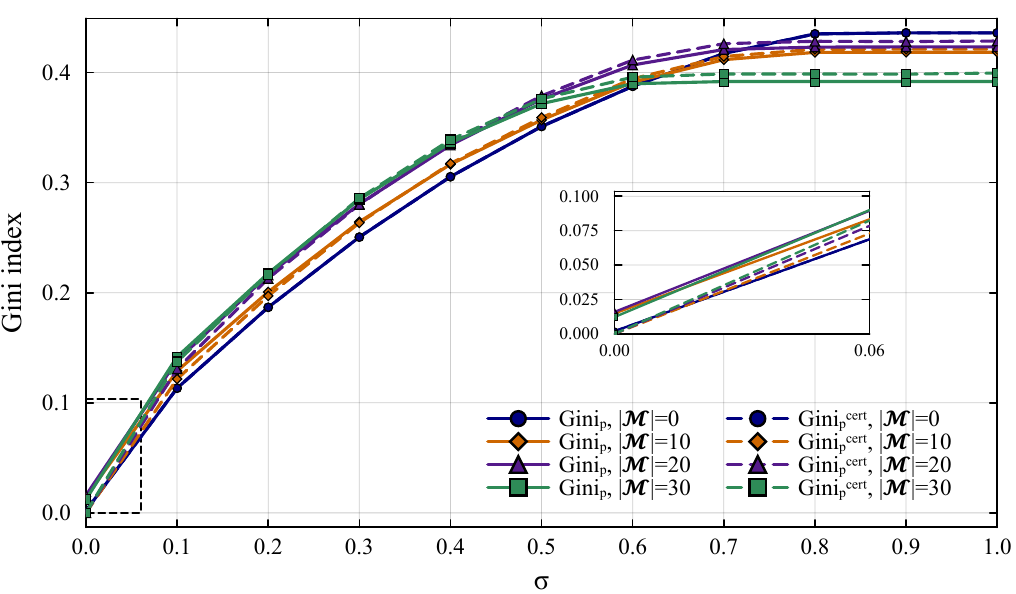}
\caption{Comparison between the certificate-based Gini index \(\mathrm{Gini}_{p}^{\mathrm{cert}}\) and the Gini index \(\mathrm{Gini}_{p}\) computed from the final polytopic DOE for different fairness settings and coordinated-set sizes.}
\label{fig:fairness_sweep_gini}
\end{figure}

We further examine whether the fairness control introduced through the ellipsoidal certificate variables, \(P_{\mathcal M}^{+}\) and \(P_{\mathcal M}^{-}\) in~\eqref{eq:directional}, is preserved in the final polytopic DOE. We use the same 3589 bus, 3588 line network and solve the coordinated DOE problem for \(|\mathcal M|=0,10,20,30\) and fairness parameter \(\sigma=0.0,0.1,\ldots,1.0\), giving \(44\) cases in total. To assess consistency, we compute \(\mathrm{Gini}_{p}^{\mathrm{cert}}\) from the optimized certificate variables and \(\mathrm{Gini}_{p}\) from the active-power ranges induced by the final polytope.
Fig.~\ref{fig:fairness_sweep_gini} shows that the certificate-based and polytope-based Gini indices increase smoothly and closely follow each other across all tested cases with a largest observed difference between \(\mathrm{Gini}_{p}^{\mathrm{cert}}\) and \(\mathrm{Gini}_{p}\) is below \(0.016\). This confirms that although the fairness constraints are enforced through certificate variables, they still control the disparity of the final communicated polytopic DOE in the tested cases.

\section{Conclusion}
\label{sec:conclusion}
This paper presented a convex, geometry-aware framework for constructing DOEs under partial coordination, combining individually valid hyperrectangular limits for non-coordinated customers with a coupled polytopal operating set for coordinated customers. The formulation admits a budgeted robust extension to incorporate forecast uncertainty in inelastic injections without relying on distributional assumptions.

To address fairness concerns about allocated flexibility across customers, we incorporated a weight-based fairness constraint with a tunable parameter that ensures that flexibility across heterogeneous customers is less dispersed. Numerical studies on the European LV test feeder confirmed that coordinating 30\% of customers increased the achievable aggregate active power injection range by approximately 25\% on average across randomized scenarios, relative to the non-coordinated baseline. Tightening fairness requirements reduced the disparity of weight-normalized allocations, at the expense of a smaller aggregate envelope, and robustification further reduced available headroom as conservatism increased. While the DOE construction is based on linearized power-flow constraints; to validate feasibility under the full nonlinear network physics, we performed adversarial AC power-flow tests, which yielded voltage and line-loading extrema consistent with operational limits.

This work opens several important directions for future study. The optimal choice of customers within the coordinating group will help analyze flexibility improvement, specifically for aggregator programs aimed at providing grid services. An extension to the DOE design for multiple cohorts of coordinating customers (each cohort within an aggregator) will help analyze the benefits under competing aggregators. While we consider weighted allocation, an extension to a price-sensitive DOE design will make the allocation more attuned to customer flexibility incentives. Finally, we plan to extend our framework to data-driven DOE design, where the network parameters or customer flexibility models are estimated using historical data.

\section*{Acknowledgement}
This work is supported by the Flemish Government and Flanders Innovation \& Entrepreneurship (VLAIO) through IMPROcap project (HBC.2022.0733), MIT Global Seed Fund, and the MIT Energy Initiative Future Energy System Center.

\balance
\bibliographystyle{IEEEtran}
\bibliography{reference} 

@techreport{masson2024trends,
  title        = {Trends in PV Applications 2024},
  author       = {Ga\u{e}tan Masson and Melodie de l'Epine and Izumi Kaizuka},
  institution  = {IEA-PVPS Task 1},
  year         = {2024},
  doi          = {10.69766/JNEW6916},
  isbn         = {978-3-907281-68-0},
  note         = {29th edition of the IEA-PVPS Trends report},
}

@techreport{iea2025global,
  title        = {Global Energy Review 2025},
  author       = {{International Energy Agency}},
  year         = {2025},
  institution  = {IEA},
  address      = {Paris},
  url          = {https://www.iea.org/reports/global-energy-review-2025}
}

@article{baran2002optimal,
  title={Optimal sizing of capacitors placed on a radial distribution system},
  author={Baran, Mesut and Wu, Felix F},
  journal={IEEE Transactions on power Delivery},
  volume={4},
  number={1},
  pages={735--743},
  year={2002},
  publisher={IEEE}
}

@article{deka2017structure,
  title={Structure learning in power distribution networks},
  author={Deka, Deepjyoti and Backhaus, Scott and Chertkov, Michael},
  journal={IEEE Transactions on Control of Network Systems},
  volume={5},
  number={3},
  pages={1061--1074},
  year={2017},
  publisher={IEEE}
}

@article{bhadoriya2025family,
  title={A Family of Convex Models to Achieve Fairness through Dispersion Control},
  author={Bhadoriya, Abhay Singh and Deka, Deepjyoti and Sundar, Kaarthik},
  journal={arXiv preprint arXiv:2510.23791},
  year={2025}
}

@standard{ieee1547.2-2023,
  title        = {IEEE Application Guide for IEEE Std 1547{\texttrademark}--2018, IEEE Standard for Interconnection and Interoperability of Distributed Energy Resources with Associated Electric Power Systems Interfaces},
  standardnumber = {IEEE 1547.2-2023},
  publisher    = {Institute of Electrical and Electronics Engineers},
  address      = {New York, NY},
  year         = {2023},
  note         = {Guide to IEEE Std 1547-2018; 291 pages}  
}

@techreport{iea2022der,
  title        = {Unlocking the Potential of Distributed Energy Resources},
  author       = {{International Energy Agency}},
  year         = {2022},
  institution  = {IEA},
  address      = {Paris},
  url          = {https://www.iea.org/reports/unlocking-the-potential-of-distributed-energy-resources}
}

@article{sundar2025parametric,
  title={A Parametric, second-order cone representable model of fairness for decision-making problems},
  author={Sundar, Kaarthik and Deka, Deepjyoti and Bent, Russell},
  journal={Optimization and Engineering},
  pages={1--16},
  year={2025},
  publisher={Springer}
}

@inproceedings{jiang2024robust,
  title={A Robust Optimization Method for Dynamic Operating Envelope Coordination in Distribution Networks},
  author={Jiang, Zhisen and Guo, Ye and Wang, Jianxiao},
  booktitle={2024 IEEE 8th Conference on Energy Internet and Energy System Integration (EI2)},
  pages={5326--5330},
  year={2024},
  organization={IEEE}
}

@article{scott2019network,
  title={Network-aware coordination of residential distributed energy resources},
  author={Scott, Paul and Gordon, Dan and Franklin, Evan and Jones, Laura and Thi{\'e}baux, Sylvie},
  journal={IEEE Transactions on Smart Grid},
  volume={10},
  number={6},
  pages={6528--6537},
  year={2019},
  publisher={IEEE}
}

@article{gao2025equitable,
  title={Equitable Active-Reactive Power Envelopes for Distributed Energy Resources in Power Distribution Systems},
  author={Gao, Yuanhai and Xu, Xiaoyuan and Yan, Zheng and Shahidehpour, Mohammad},
  journal={IEEE Transactions on Smart Grid},
  year={2025},
  publisher={IEEE}
}

@inproceedings{hashmi2023robust,
  title={Robust dynamic operating envelopes for flexibility operation using only local voltage measurement},
  author={Hashmi, Md Umar and Van Hertem, Dirk},
  booktitle={2023 International Conference on Smart Energy Systems and Technologies (SEST)},
  pages={1--6},
  year={2023},
  organization={IEEE}
}

@article{kekatos2015voltage,
  title={Voltage regulation algorithms for multiphase power distribution grids},
  author={Kekatos, Vassilis and Zhang, Liang and Giannakis, Georgios B and Baldick, Ross},
  journal={IEEE Transactions on Power Systems},
  volume={31},
  number={5},
  pages={3913--3923},
  year={2015},
  publisher={IEEE}
}

@misc{ieee2015european,
  title        = {{IEEE} {PES} {T}est {F}eeder: {E}uropean Low Voltage Test Feeder},
  year         = {2015},
  howpublished = {\url{https://cmte.ieee.org/pes-testfeeders/resources/}},
  note         = {[Online]},
  organization = {IEEE Power and Energy Society},
}

@article{jain1984quantitative,
  title={A quantitative measure of fairness and discrimination},
  author={Jain, Rajendra K and Chiu, Dah-Ming W and Hawe, William R and others},
  journal={Eastern Research Laboratory, Digital Equipment Corporation, Hudson, MA},
  volume={21},
  number={1},
  pages={2022--2023},
  year={1984}
}

@article{gini1936measure,
  title={On the measure of concentration with special reference to income and statistics, Colorado College Publication},
  author={Gini, Corrado},
  journal={General series},
  volume={208},
  number={1},
  year={1936}
}

@inproceedings{rubbers2025fairness,
  title={Fairness for distribution network hosting capacity},
  author={Rubbers, Olivia and Kerckhove, Sari and Hashmi, Md Umar and Van Hertem, Dirk},
  booktitle={2025 IEEE PES Innovative Smart Grid Technologies Conference Europe (ISGT Europe)},
  pages={1--5},
  year={2025},
  organization={IEEE}
}

@article{liu2021project,
  title={Project {EDGE}-Deliverable 1.1: Operating Envelopes Calculation Architecture},
  author={Liu, Michael Z and Ochoa, Luis Nando},
  year={2021}
}

@article{liu2021grid,
  title={Grid and market services from the edge: Using operating envelopes to unlock network-aware bottom-up flexibility},
  author={Liu, Michael Z and Ochoa, Luis Nando and Riaz, Shariq and Mancarella, Pierluigi and Ting, Tian and San, Jack and Theunissen, John},
  journal={IEEE Power and Energy Magazine},
  volume={19},
  number={4},
  pages={52--62},
  year={2021},
  publisher={IEEE}
}

@article{hashmi2022can,
  title={Can locational disparity of prosumer energy optimization due to inverter rules be limited?},
  author={Hashmi, Md Umar and Deka, Deepjyoti and Bu{\v{s}}i{\'c}, Ana and Van Hertem, Dirk},
  journal={IEEE Transactions on Power Systems},
  volume={38},
  number={6},
  pages={5726--5739},
  year={2022},
  publisher={IEEE}
}

@article{alam2023allocation,
  title={Allocation of dynamic operating envelopes in distribution networks: Technical and equitable perspectives},
  author={Alam, Mollah Rezaul and Nguyen, Phuong TH and Naranpanawe, Lakshitha and Saha, Tapan K and Lankeshwara, Gayan},
  journal={IEEE Transactions on Sustainable Energy},
  volume={15},
  number={1},
  pages={173--186},
  year={2023},
  publisher={IEEE}
}

@article{givisiez2024accelerating,
  title={Accelerating the Implementation of Operating Envelopes Across {A}ustralia--Milestone 4},
  author={Arthur Gonçalves Givisiez and Luis(Nando) Ochoa},
  year={2024}
}

@misc{aemo_projectedge,
  title        = {Project EDGE},
  author       = {{Australian Energy Market Operator}},
  year         = {2022},
  howpublished = {\url{https://aemo.com.au/en/initiatives/major-programs/nem-distributed-energy-resources-der-program}},
}

@techreport{esb2021flexibleaccess,
  title        = {Flexible Access to the Distribution System for Renewable Generators -- Pilot Project 4B Final Report},
  author       = {{ESB Networks}},
  year         = {2021},
  
}

@inproceedings{foote2013orkney,
  author       = {C. Foote and R. Johnston and F. Watson and R. Currie and D. Macleman and A. Urquhart},
  title        = {Second Generation Active Network Management on Orkney},
  booktitle    = {22nd International Conference and Exhibition on Electricity Distribution (CIRED 2013)},
  year         = {2013},
  pages        = {0659},
  doi          = {10.1049/cp.2013.0864},
  url          = {https://digital-library.theiet.org/doi/abs/10.1049/cp.2013.0864}
}

@article{yi2022fair,
  title={Fair operating envelopes under uncertainty using chance constrained optimal power flow},
  author={Yi, Yu and Verbi{\v{c}}, Gregor},
  journal={Electric Power Systems Research},
  volume={213},
  pages={108465},
  year={2022},
  publisher={Elsevier}
}

@article{liu2023robust,
  title={Robust dynamic operating envelopes for {DER} integration in unbalanced distribution networks},
  author={Liu, Bin and Braslavsky, Julio H},
  journal={IEEE Transactions on Power Systems},
  volume={39},
  number={2},
  pages={3921--3936},
  year={2023},
  publisher={IEEE}
}

@article{lankeshwara2023time,
  title={Time-varying operating regions of end-users and feeders in low-voltage distribution networks},
  author={Lankeshwara, Gayan and Sharma, Rahul and Yan, Ruifeng and Saha, Tapan K and Milanovi{\'c}, Jovica V},
  journal={IEEE Transactions on Power Systems},
  year={2023},
  publisher={IEEE}
}

@article{BJARGHOV2024100154,
    title = {Enhancing grid hosting capacity with coordinated non-firm connections in industrial energy communities},
    journal = {Smart Energy},
    volume = {15},
    pages = {100154},
    year = {2024},
    issn = {2666-9552},
    doi = {https://doi.org/10.1016/j.segy.2024.100154},
    url = {https://www.sciencedirect.com/science/article/pii/S2666955224000248},
    author = {Sigurd Bjarghov and Sverre {Stefanussen Foslie} and Magnus Askeland and Rubi Rana and Henning Taxt},
}

@Manual{volesti,
    title = {volesti: Volume Approximation and Sampling of Convex Polytopes},
    author = {Vissarion Fisikopoulos and Apostolos Chalkis},
    year = {2025},
    note = {R package version 1.1.2-9},
    url = {https://CRAN.R-project.org/package=volesti},
    doi = {10.32614/CRAN.package.volesti},
}

@article{ahmadi2014linear,
  title={Linear current flow equations with application to distribution systems reconfiguration},
  author={Ahmadi, Hamed and Mart{\i}, Jos{\'e} R},
  journal={IEEE Transactions on Power Systems},
  volume={30},
  number={4},
  pages={2073--2080},
  year={2014},
  publisher={IEEE}
}

@book{boyd2004convex,
  title={Convex optimization},
  author={Boyd, Stephen P and Vandenberghe, Lieven},
  year={2004},
  publisher={Cambridge university press}
}

@article{azim2024dynamic,
  title={Dynamic Operating Envelope-Integrated Cooperative Energy Trading: A Fair Allocation Approach},
  author={Azim, M Imran and Hoque, Md Murshadul and Khorasany, Mohsen and Razzaghi, Reza and Jalili, Mahdi and Hill, David J},
  journal={IEEE Transactions on Power Systems},
  year={2024},
  publisher={IEEE}
}

@inproceedings{russell2023robust,
  title={Robust Operating Envelopes with Phase Unbalance Constraints in Unbalanced Three-Phase Networks},
  author={Russell, James Stanley and Scott, Paul and Iria, Jos{\'e}},
  booktitle={2023 IEEE PES Innovative Smart Grid Technologies-Asia (ISGT Asia)},
  pages={1--5},
  year={2023},
  organization={IEEE}
}

@inproceedings{gan2014convex,
  title={Convex relaxations and linear approximation for optimal power flow in multiphase radial networks},
  author={Gan, Lingwen and Low, Steven H},
  booktitle={2014 power systems computation conference},
  pages={1--9},
  year={2014},
  organization={IEEE}
}

@article{bertsimas2004price,
  title={The price of robustness},
  author={Bertsimas, Dimitris and Sim, Melvyn},
  journal={Operations research},
  volume={52},
  number={1},
  pages={35--53},
  year={2004},
  publisher={Informs}
}

@article{riaz2021modelling,
  title={Modelling and characterisation of flexibility from distributed energy resources},
  author={Riaz, Shariq and Mancarella, Pierluigi},
  journal={IEEE transactions on power systems},
  volume={37},
  number={1},
  pages={38--50},
  year={2021},
  publisher={IEEE}
}

@article{song2025distribution,
  title={Distribution dynamic operating security space characterization for aggregation and congestion capacity evaluation of virtual power plant},
  author={Song, Meng and Xu, Zekai and Gao, Ciwei and Yan, Mingyu and Hu, Qinran},
  journal={IEEE Transactions on Smart Grid},
  year={2025},
  publisher={IEEE}
}

@article{mahmoodi2023capacity,
  title={Der capacity assessment of active distribution systems using dynamic operating envelopes},
  author={Mahmoodi, Masoume and Blackhall, Lachlan and RA, S Mahdi Noori and Attarha, Ahmad and Weise, Ben and Bhardwaj, Abhishek},
  journal={IEEE Transactions on Smart Grid},
  year={2023},
  publisher={IEEE}
}

@article{jiang2025bargaining,
  title={Bargaining-based approach for dynamic operating envelope allocation in distribution networks},
  author={Jiang, Zhisen and Guo, Ye},
  journal={IEEE Transactions on Smart Grid},
  year={2025},
  publisher={IEEE}
}

@article{petrou2021ensuring,
  title={Ensuring distribution network integrity using dynamic operating limits for prosumers},
  author={Petrou, Kyriacos and Procopiou, Andreas T and Gutierrez-Lagos, Luis and Liu, Michael Z and Ochoa, Luis F and Langstaff, Tom and Theunissen, John M},
  journal={IEEE Transactions on Smart Grid},
  volume={12},
  number={5},
  pages={3877--3888},
  year={2021},
  publisher={IEEE}
}

@article{hashminetwork,
title = {Network aware energy market participation of industrial flexibility},
journal = {Sustainable Energy, Grids and Networks},
pages = {102460},
year = {2026},
issn = {2352-4677},
doi = {https://doi.org/10.1016/j.segan.2026.102460},
url = {https://www.sciencedirect.com/science/article/pii/S2352467726003425},
author = {Md Umar Hashmi and Carlo Manna and Dirk {Van Hertem}},
}

@inproceedings{fani2024impact,
  title={Impact of dynamic operating envelopes on distribution network hosting capacity for electric vehicles},
  author={Fani, Hossein and Hashmi, Md Umar and Palacios-Garcia, Emilio J and Deconinck, Geert},
  booktitle={IET Conference Proceedings CP876},
  volume={2024},
  number={5},
  pages={180--185},
  year={2024},
  organization={IET}
}

@article{hoque2023dynamic,
  title={Dynamic operating envelope-based local energy market for prosumers with electric vehicles},
  author={Hoque, Md Murshadul and Khorasany, Mohsen and Azim, M Imran and Razzaghi, Reza and Jalili, Mahdi},
  journal={IEEE Transactions on Smart Grid},
  volume={15},
  number={2},
  pages={1712--1724},
  year={2023},
  publisher={IEEE}
}

@article{alahmed2024decentralized,
  title={A decentralized market mechanism for energy communities under operating envelopes},
  author={Alahmed, Ahmed S and Cavraro, Guido and Bernstein, Andrey and Tong, Lang},
  journal={IEEE Transactions on Control of Network Systems},
  volume={12},
  number={1},
  pages={313--324},
  year={2024},
  publisher={IEEE}
}

@techreport{SAPN_AusNet_FlexibleExports,
  author      = {{SA Power Networks}},
  title       = {{Flexible Exports for Solar PV Trial: Final Report}},
  institution = {{Australian Renewable Energy Agency}},
  year        = {2024},
  month       = jan,
  url         = {https://arena.gov.au/knowledge-bank/sa-power-networks-flexible-exports-for-solar-pv-trial-final-report/},
  note        = {Accessed: 2026-06-23}
}

@techreport{AER_FlexibleExportLimits,
  author      = {{Australian Energy Regulator}},
  title       = {{Response to Flexible Export Limits Consultation}},
  institution = {{Australian Energy Regulator}},
  year        = {2023},
  month       = jul,
  url         = {https://www.aer.gov.au/documents/aer-response-flexible-export-limits-consultation-july-2023},
  note        = {Accessed: 2026-06-23}
}

@misc{EnergyQueenslandDynamicConnections,
  author       = {{Energex}},
  title        = {{About Dynamic Connections}},
  year         = {2025},
  howpublished = {Online},
  url          = {https://www.energex.com.au/our-services/connections/residential-and-commercial-connections/solar-connections-and-other-technologies/dynamic-connections-for-energy-exports/about-dynamic-connections},
  note         = {Accessed: 2026-06-23}
}

@article{attarha2021network,
  title={Network-secure envelopes enabling reliable DER bidding in energy and reserve markets},
  author={Attarha, Ahmad and RA, S Mahdi Noori and Scott, Paul and Thi{\'e}baux, Sylvie},
  journal={IEEE Transactions on Smart Grid},
  volume={13},
  number={3},
  pages={2050--2062},
  year={2021},
  publisher={IEEE}
}

@article{espinosa2015dissemination,
  title={Dissemination Document “Low Voltage Networks Models and Low Carbon Technology Profiles”},
  author={Espinosa, Alejandro Navarro},
  year={2015}
}

@article{geth2025four,
  title={Four-Wire-OPF-Based DOE Quantification Incorporating Volt-var/Watt Response},
  author={Geth, Frederik and Lucas, Kurt and Clark, Jordan and Nimalsiri, Nanduni and Heidari, Rahmat},
  year={2025},
  publisher={GridQube and CSIRO}
}

@article{gerdroodbari2022dynamic,
  title={Dynamic PQ operating envelopes for prosumers in distribution networks},
  author={Gerdroodbari, Yasin Zabihinia and Khorasany, Mohsen and Razzaghi, Reza},
  journal={Applied Energy},
  volume={325},
  pages={119757},
  year={2022},
  publisher={Elsevier}
}

@article{Kumarawadu2025SmartMeter,
  title={Smart meter data-driven dynamic operating envelopes for DERs},
  author={Kumarawadu, Achala and Azim, M Imran and Khorasany, Mohsen and Razzaghi, Reza and Heidari, Rahmat},
  journal={Applied Energy},
  volume={384},
  pages={125469},
  year={2025},
  publisher={Elsevier}
}
\end{document}